\documentclass[review,numbers,sort&compress]{elsarticle}

\usepackage[utf8]{inputenc}
\usepackage{enumerate}
\usepackage{amsmath}
\usepackage{amsfonts}
\usepackage{enumitem}
\usepackage{etoolbox}
\usepackage{multirow}
\usepackage{natbib}
\usepackage{makecell}
\usepackage{setspace}

\usepackage{lineno,hyperref}
\usepackage[letterpaper,%
            left=1in,right=1.75in,top=1in,bottom=1in,%
            footskip=.25in]{geometry}

\journal{Elsevier}

\usepackage{caption}
\usepackage{amsmath}
\usepackage{graphicx}
\usepackage{subcaption}
\usepackage{graphicx}
\usepackage{todonotes}
\usepackage{nomencl}
\usepackage{bm}
\usepackage[final]{changes}
\usepackage{xcolor}
\usepackage{etoolbox}
\usepackage{comment}
\graphicspath{{./}}
\renewcommand\nomgroup[1]{%
  \item[\itshape
  \ifstrequal{#1}{A}{Symbols}{%
  \ifstrequal{#1}{B}{Roman Letters}{%
  \ifstrequal{#1}{C}{Greek Letters}{%
  \ifstrequal{#1}{D}{Abbreviations}{}}}}%
]}

\begin{document}

\begin{frontmatter}
\title{
    A features-embedded-learning immersed boundary model for large-eddy simulation of turbulent flows with complex boundaries
}

  \author[cas,ucas]{Zhideng Zhou}
  \author[cas,ucas]{Fengshun Zhang}
  \author[cas,ucas]{Xiaolei Yang\corref{mycor}}
  \ead{xyang@imech.ac.cn}
  \cortext[mycor]{Corresponding author}

  \address[cas]{The State Key Laboratory of Nonlinear Mechanics, Institute of Mechanics, Chinese Academy of Sciences,
Beijing 100190, China}
  \address[ucas]{School of Engineering Sciences, University of Chinese Academy of Sciences, Beijing 100049, China}

\begin{abstract}
%
%
The hybrid wall-modeled large-eddy simulation (WMLES) and immersed boundary (IB) method offers significant flexibility for simulating high Reynolds number flows involving complex boundaries. However, the approximate boundary conditions (e.g., the wall shear stress boundary condition) developed for body-fitted grids in the literature are not directly applicable to IB methods. In this work, we propose a features-embedded-learning-IB (FEL-IB) wall model to approximate IB boundaries in the hybrid WMLES-IB method, in which the velocity at the IB node (a grid node located in the fluid that has at least one neighbor in the solid) is reconstructed using the power law of the wall, and the momentum flux at the interface between the IB nodes and the fluid nodes is approximated using a neural network model. The neural network model for momentum flux is first pretrained using high-fidelity simulation data of the flow over periodic hills and the logarithmic law, and then learned in the WMLES environment using the ensemble Kalman method. The proposed model is evaluated using two challenging cases: the flow over a body of revolution and the DARPA Suboff submarine model. For the first case, good agreement with the reference data is obtained for the vertical profiles of the streamwise velocity. For the flow over the DARPA Suboff submarine model, WMLES cases with two Reynolds numbers and two grid resolutions are carried out. Overall good \emph{a posteriori} performance is observed for predicting the mean and root mean square of velocity profiles at various streamwise locations, as well as the skin-friction and pressure coefficients.
\end{abstract}

\end{frontmatter}


\section{Introduction}\label{sec:Introduction}
The immersed boundary (IB) method is an essential tool for simulating flows with geometric complex or moving boundaries~\cite{Mittal_Iaccarino_ARFM_2005, Verzicco_ARFM_2023}. While wall models enable practical large-eddy simulations (LES) of high Reynolds number flows by avoiding direct resolution of near-wall turbulence~\cite{Bose_Park_ARFM_2018, Choi_Moin_PoF_2012, Yang_Griffin_PoF_2021}, their integration with IB methods remains challenging. The unavoidable mismatch between the background grid and the boundaries in IB formulations prevents direct application of conventional wall models, particularly for imposing accurate wall shear stress boundary conditions. To address this problem, we propose a machine learning-based approach for modeling immersed boundaries within a sharp-interface curvilinear IB framework.

In recent years, the machine learning method has become a promising tool for developing LES wall models for high Reynolds number wall-bounded turbulent flows. The predictive capability of a data-driven model is often constrained by the training data, limiting its applications to different flow configurations. Various strategies have been developed in the literature. One way is incorporating the known physics into model training. In the work by Yang \emph{et al.}~\cite{Yang_etal_PRF_2019}, a feedforward neural network (FNN) model was constructed to compute the wall shear stress using the input features normalized by the logarithmic law~\cite{Deardorff_JFM_1970}, and successfully applied to the wall-modelled LES (WMLES) of turbulent channel flows at various Reynolds numbers and spanwise rotating turbulent channel flows~\cite{Huang_etal_PoF_2019}. In the work by Lee~\emph{et al.}~\cite{Lee_etal_AST_2023}, the Fukagata-Iwamoto-Kasagi identity~\cite{FIK_identity_PoF_2002} was used to extract input features for determining the wall shear stress. The other way is employing dataset from various flow regimes for model training. Zangeneh~\cite{Zangeneh_PoF_2021} used the DNS (direct numerical simulation) data from the zero pressure gradient turbulent flow over a flat plate and supersonic flow around an expansion-compression corner to train a random forest regression model for wall shear stress and heat flux. Dupuy \emph{et al.}~\cite{Dupuy_etal_JCP_2023} used the filtered high-fidelity data from the turbulent channel flows and the separated flows in a three-dimensional diffuser and over a backward-facing step to train a FNN model of wall shear stress. In our previous work~\cite{Zhou_etal_PoF_2023}, the wall-resolved LES (WRLES) data of the periodic hill separated flows and the logarithmic law of the wall are integrated to train a FNN model for predicting wall shear stress. For separated flows, we proposed to employ the flow features at three off-wall grid nodes for near-wall flow modelling to solve the issue that the wall shear stress cannot be uniquely determined using the wall-tangential velocity at one single off-wall grid node~\cite{Zhou_He_Yang_PRF_2021}. In the work by Lozano-Durán \& Bae~\cite{Lozano_Bae_JFM_2023}, the building-block-flow wall model, which integrates submodels trained using various canonical flows, is proposed for predicting complex flows. 
%
%

The data for model training are often from the DNS and WRLES, different from the WMLES environment, causing suboptimal \emph{a posteriori} performance. Various strategies have been proposed to solve the issue.
The so-called model-consistent training strategy was employed by Bae \& Koumoutsakos~\cite{bae2022scientific}, who used the multi-agent reinforcement learning (MARL) to train wall eddy-viscosity models for turbulent channel flows in the WMLES environment. This MARL method was later applied to train wall models for the flow over periodic hills.
In the work by Lozano-Durán \& Bae~\cite{Lozano_Bae_JFM_2023}, the model-consistent training was achieved by using WMLES data obtained with an optimised SGS model as the training dataset, instead of the DNS data~\cite{Lozano_Duran_Bae_2020}. In our recent work~\cite{Zhou_etal_JFM_2025}, a features-embedded-learning (FEL) wall model was proposed for separated flows, which comprises a FNN model for estimating wall shear stress and another FNN model for modelling the eddy viscosity at the first off-wall grid nodes. The former is trained using the WRLES data of flow over periodic hills that contains various flow features along the streamwise direction (e.g., flow separation, recirculation, reattachment, recovery and strong acceleration), and the law of the wall~\cite{Zhou_etal_PoF_2023}. For the latter, a modified mixing length model is proposed with the coefficient trained in the WMLES environment using the ensemble Kalman method, which ensures the model consistency. Successful applications to separated flows with different configurations, grid resolutions and Reynolds numbers were demonstrated.


Studies on data-driven wall models have been focused on the body-fitted grids. Since the immersed boundaries are not discretized by the background grid for solving the fluid flow, the wall shear stress boundary conditions cannot be directly applied in IB methods. 
%
%
For the developments of wall models for the diffuse-interface IB method, the readers can refer to these works~\cite{Shi_2019_WMLES, Ma_etal_IJHFF_2021, Du_etal_IJNMF_2022, Yan_etal_PoF_2024}.
This work focuses on the development of data-driven wall models for the sharp-interface IB method. In this type of IB method, the velocity at the IB nodes (in this work, the IB node is defined as the fluid node with at least one neighbor in the solid) is reconstructed from the near-wall flow and the velocity at the wall to provide boundary conditions for the outer flow simulations. Various near-wall modelling approaches have been developed for the sharp-interface IB method. 
%
In the work of Tessicini \textit{et al.}~\cite{Tessicini_etal_2002}, an equilibrium wall model~\cite{Wang_Moin_PoF_2002} formulated by solving the simplified boundary layer equations in the body-fitted grids was implemented in IB method to reconstruct the flow velocity between the second off-wall grid points and the wall for a hydrofoil trailing edge. 
Choi \textit{et al.}~\cite{Choi_etal_JCP_2007} used the power law to construct the tangential velocity near the immersed surface. Roman \textit{et al.}~\cite{Roman_etal_PoF_2009} proposed a two-step wall model in IB method, while the velocity at the first off-wall grid is reconstructed by the logarithmic law and then the wall shear stress at the immersed surface is reconstructed by imposing a Reynolds-averaged Navier-Stokes-like (RANS) eddy viscosity. 
Tamaki \& Kawai~\cite{Tamaki_Kawai_PRF_2021} proposed an approach to realize WMLESs of fully developed flat-plate turbulent boundary layers with the IB method by considering the shear-stress balance in the near-wall region. The shear-stress balance is maintained by employing a partial-slip wall-boundary condition to remedy the underresolution in the near-wall region and using an algebraic wall function to model the turbulent shear stress to compensate for the reduction of velocity gradient due to the partial-slip velocity boundary condition.
One significant issue with the above studies is that the wall models based on the equilibrium hypothesis 
are in general not applicable to flows with geometric complex boundaries.
%

\begin{figure}[!ht]
\centering{\includegraphics[width=0.9\textwidth]{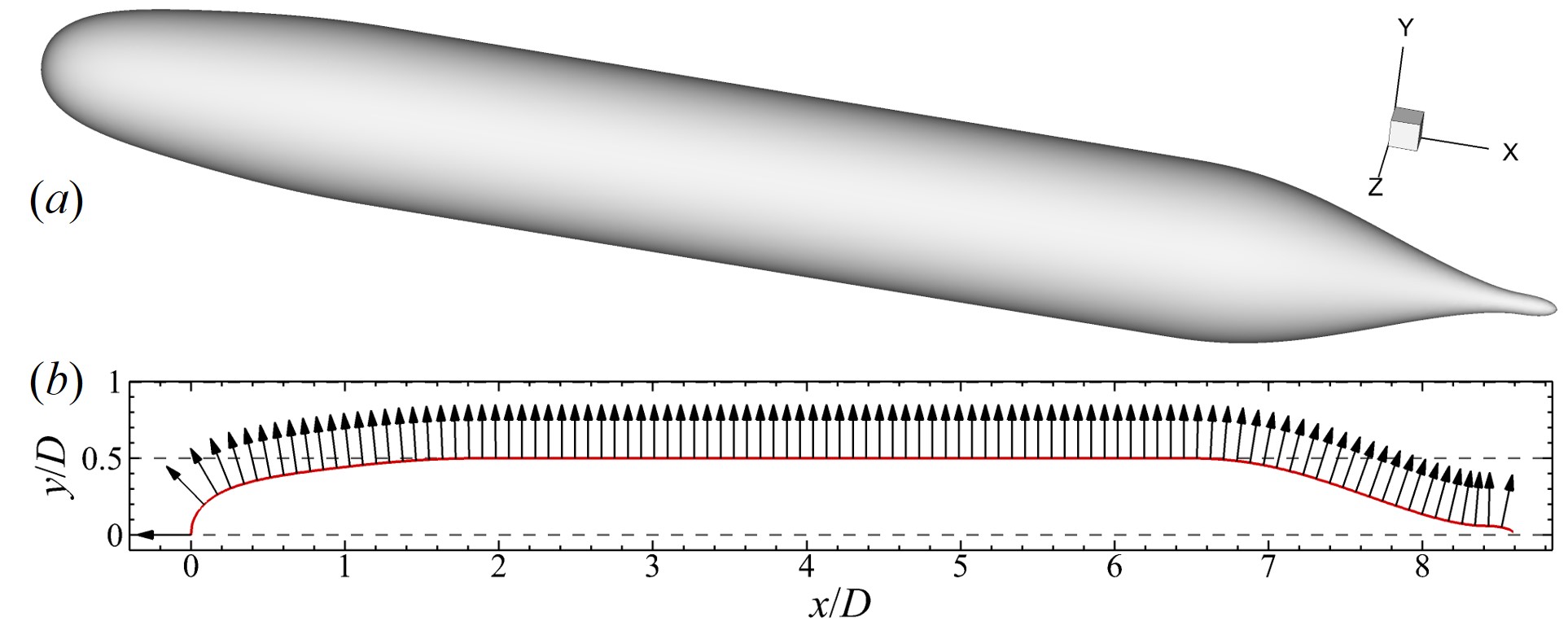}}
  \caption{{\color{black} Geometry (a) and profile (a) of the bare hull of the DARPA Suboff model.}}
\label{fig:geometry}
\end{figure}

The objective of this work is to develop a machine-learning-augmented wall model for approximating the IB boundaries for a sharp-interface IB method. 
The key test case of the proposed model is the flow over 
the Defense Advance Research Projects Agency (DARPA) Suboff submarine model~\cite{Groves_etal_1989_techreport} (figure~\ref{fig:geometry}),
a notional geometry for investigating the hydrodynamics of underwater vehicles. The flow over the Suboff model features transition (at the streamlined forebody), boundary layer development (at the parallel middle body), thickening and separation (at the stern with contraction)~\cite{Huang_1994_measurements}, establishing it as a challenging test case for LES and RANS models~\cite{Posa_Balaras_JFM_2016, Kumar_Mahesh_JFM_2018, Liu_etal_JCP_2023, Posa_Balaras_JFM_2020, Zhou_etal_AIAA_2020, Shi_2019_WMLES, Chen_etal_JoH_2023}.

The rest of the paper is organized as follows. In Sec. \ref{sec:Methodology}, the flow solver based on curvilinear IB methods and the FEL-IB shear stress model are introduced. In Sec. \ref{sec:ML_strategy}, the embedded training of the FEL-IB model is described. 
The test cases and corresponding results are then presented in Sec. \ref{sec:Test_cases}. At last, the conclusions are drawn in Sec. \ref{sec:Conclusion}.

\section{Methodology}\label{sec:Methodology}

In this section, we first introduce the flow solver based on curvilinear immersed boundary methods. Then the FEL-IB model is proposed to approximate the turbulent shear stress and modify the momentum flux on IB boundaries for the sharp-interface IB method.

\subsection{Flow solver based on curvilinear immersed boundary methods}\label{sec:2.1}

We employ the Virtual Flow Simulator (VFS-Wind)~\cite{Yang_etal_WE_2015, Yang_Sotiropoulos_WE_2018, Zhou_Wu_Yang_PoF_2021, Zhou_etal_TAML_2021} code for the LES of turbulent flows over the complex boundaries.
The governing equations are the spatially filtered incompressible Navier-Stokes equations in non-orthogonal, generalized curvilinear coordinates as follows,
\begin{equation}
\left\{
  \begin{aligned}
    J \frac{ \partial U^{j} }{\partial \xi^{j}} &= 0, \\
    \frac{1}{J} \frac{\partial U^{i}}{\partial t} &= \frac{\xi^{i}_{l}}{J} \left( -\frac{\partial}{\partial \xi^{j}} (U^{j}u_{l}) - \frac{1}{\rho} \frac{\partial}{\partial \xi^{j}} ( \frac{\xi^{j}_{l} p}{J} ) + \frac{\mu}{\rho} \frac{\partial}{\partial \xi^{j}}( \frac{g^{jk}}{J} \frac{\partial u_{l}}{\partial \xi^{k}} ) - \frac{1}{\rho} \frac{\partial \tau_{lj}}{\partial \xi^{j}} + f_l \right),
  \end{aligned}
\right.
\label{eq_1}
\end{equation}
where $x_{i}$ and $\xi^{i}$ are the Cartesian and curvilinear coordinates, respectively, $\xi^{i}_{l} = \partial \xi^{i} / \partial x_{l}$ are the transformation metrics, $J$ is the Jacobian of the geometric transformation, $u_{i}$ is the $i$-th component of the velocity vector in Cartesian coordinates, $U^{i} = (\xi^{i}_{m} / J) u_{m}$ is the contravariant volume flux, $g^{jk} = \xi^{j}_{l} \xi^{k}_{l}$ are the components of the contravariant metric tensor, $\rho$ is the fluid density, $\mu$ is the dynamic viscosity, and $p$ is the pressure. In the momentum equation, $\tau_{ij}$ represents the anisotropic part of the SGS stress tensor, which is modeled using the dynamic Smagorinsky model~\citep{Germano_etal_PoF_1991, Lilly_PoF_1992}:
\begin{equation}
\tau_{ij} - \frac{1}{3} \tau_{kk} \delta_{ij} = - 2 \nu_{t} \overline{S_{ij}}, \quad \nu_{t} = (C_S \Delta)^2 |\overline{S}|,
\label{eq_2}
\end{equation}
where $\nu_t$ is the eddy viscosity, the overbar denotes the grid filtering, $\overline{S_{ij}} = \frac{1}{2} \left( \frac{\partial \overline{u_i}}{\partial x_j} + \frac{\partial \overline{u_j}}{\partial x_i} \right)$ is the filtered strain-rate tensor, $|\overline{S}| = \sqrt{2\overline{S_{ij}}\,\overline{S_{ij}}}$, $\Delta = J^{-1/3}$ is the filter size and $J^{-1}$ is the cell volume. The model coefficient $C_S$ is calculated to minimize the mean square error between the resolved stress at the grid filter and the test filter,
\begin{equation}
C_S^2 = \frac{\left\langle L_{ij}M_{ij} \right\rangle}{\left\langle M_{kl}M_{kl} \right\rangle},
\label{eq_CS}
\end{equation}
where $L_{ij} = \widehat{\overline{u_i u_j}} - \widehat{\overline{u_i}}\,\widehat{\overline{u_j}}$, $M_{ij} = 2\Delta^2 \widehat{\overline{S_{ij}}|\overline{S}|} - 2\widehat{\Delta}^2 \overline{S_{ij}} |\widehat{\overline{S}}|$, $\widehat{\Delta}$ is the size of test filter, the hat denotes the test filtering which involves the 27 grid nodes surrounding a given grid node in three dimensions.

The governing equations are spatially discretized using a second-order accurate central difference scheme, and integrated in time using the second-order accurate fractional step method. An algebraic multigrid acceleration along with generalized minimal residual method is used to solve the pressure Poisson equation. A matrix-free Newton-Krylov method is used for solving the discretized momentum equation.

\begin{figure}[!ht]
\centering{\includegraphics[width=0.4\textwidth]{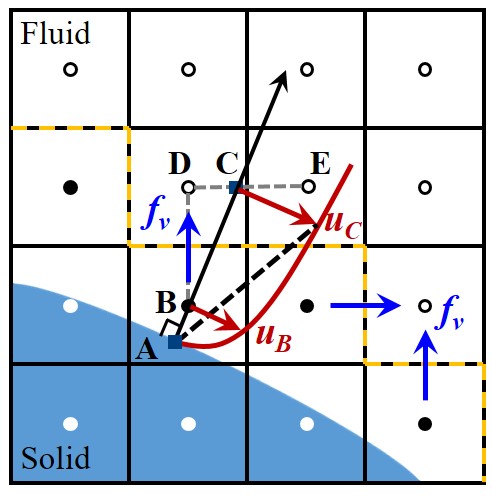}}
  \caption{The schematic of solid boundary and Cartesian background grid nodes in the WMLES using the sharp-interface IB method.}
\label{fig:IB_node}
\end{figure}

The sharp-interface IB method~\cite{Gilmanov_Sotiropoulos_JCP_2005, Ge_Sotiropoulos_JCP_2007} is employed to take into account the effect of the complex solid boundary on the surrounding flow field, in which the governing equations are discretized and solved on non-body-fitted grids. 
The geometry of the solid body is discretized using unstructured triangular meshes and superposed on the Cartesian background grid. The background grids in general do not coincide with the triangular meshes, that the boundary conditions cannot be applied directly. According to the location with respect to the solid boundary of the body, the background grid nodes are classified into solid nodes, fluid nodes and IB nodes, as shown in figure~\ref{fig:IB_node}. The solid nodes (denoted as white solid circle) that fall inside the body are 
blanked out from the simulation. The grid nodes that are located in the fluid but with at least one neighbor in the solid are identified as the IB nodes (denoted as black solid circle), where the boundary conditions are applied. The other grid nodes (denoted as black hollow circle) are identified as fluid nodes, which are used for the practical simulation. The interfaces between the IB nodes and fluid nodes (denoted as yellow dashed line) are called as the IB boundaries, on which the effect of the solid boundary on the surrounding flows is imposed as the turbulent shear stress.

%
For the DNS and WRLES cases, the velocity at the IB nodes can be simply calculated from the velocities at the fluid nodes and the wall using the linear interpolation. For WMLES case with the coarse grids, there are two modeling issues we need to deal with: (i) the linear interpolation for the velocity at the IB nodes will lead to significant error of flow prediction, because the IB nodes do not fall inside the viscous sublayer; (ii) the geometric boundary does not coincide with the computational grid, making it unable to directly give the wall-tangential velocity distribution at the wall-normal direction, so it is difficult to modify the momentum flux on the wall in the tangential direction.

To address the issue (i), a wall model is needed to reconstruct the velocity at IB nodes. In this work, we employ the Werner-Wengle (WW) model~\cite{Werner_Wengle_1993}, which computes the wall shear stress based on the linear profile and the power law for the viscous sublayer and above, respectively,
\begin{equation}
{u^+} = \left\{
  \begin{array}{ll}
    {y^+,}         &  \quad  {y^+ \le y_0^+,}   \\
    {A (y^+)^B,}   &  \quad   {y^+ > y_0^+,}    \\
  \end{array}
\right.
\label{eq_WW_model}
\end{equation}
where $A=8.3$, $b=1/7$, $y_0^+ = A^{1/(1-B)}$, $u^+ = u/u_\tau$, $y^+ = y_n u_\tau / \nu$, $y_n$ is the wall-normal distance, $u$ is the fluid velocity tangential to the solid boundary.
Based on the WW model, the friction velocity $u_\tau$ can be computed from the wall-tangential velocity $u_{t,C}$ and wall-normal distance $\Delta y_C$ at the fluid node ``C'', as shown in figure~\ref{fig:IB_node}. Then the $u_\tau$ is used to reconstruct the wall-tangential velocity at the IB node ``B'', while the wall-normal velocity is still computed using the linear interpolation.

To address the issue (ii), a neural network model will be proposed in section \ref{sec:2.2} to approximate the turbulent shear stress and modify the momentum flux on the IB boundaries.

\subsection{The FEL-IB shear stress model}\label{sec:2.2}


\subsubsection{A modified model for near-wall momentum flux based on neural network}\label{sec:2.2.2}


To approximate the turbulent shear stress and modify momentum flux on the IB boundaries, a feedforward neural network (FNN) model is first applied to predict the wall shear stress using the multi-point flow quantities in the near-wall region,
\begin{equation}
\tau_{w, \text{NN}} = f_{\text{NN}} \left( X_I \right),
\label{eq_FNN_tauw}
\end{equation}
where $X_I$ denotes the input features for the FNN model, which includes the various flow quantities at three different off-wall locations in the wall-normal direction.
These flow quantities are calculated from the interpolation of fluid nodes. The details of the FNN model are introduced in section~\ref{subsec:tau_w}. The predicted wall shear stress cannot be used directly as the boundary condition due to the inconsistency between the grid nodes and the solid boundary. Instead, it is applied to the flow solver through the form of eddy viscosity at the wall~\cite{bae2022scientific},
\begin{equation}
\left. \nu_t \right|_w = \left. \frac{d U_t}{dn} \right|_w^{-1} \frac{\tau_{w, \text{NN}}}{\rho} - \nu, \quad {\left. \frac{d U_t}{dn} \right|_w} = \frac{u_{t,C}}{\Delta y_C},
\label{eq_FNN_nut}
\end{equation}
where $\nu_t$ is the eddy viscosity and the subscript $w$ indicates values evaluated at the wall. 
The eddy viscosity at the IB nodes can be computed using the linear interpolation,
\begin{equation}
\left. {{\nu _t}} \right|_B = \frac{\Delta {y_B} \cdot \left( \left. {{\nu_t}} \right|_C - \left. {{\nu _t}} \right|_w \right)}{\Delta {y_C}} + {\left. \nu_t \right|_w}.
\label{eq_FNN_nut_IB}
\end{equation}
Then the momentum flux at the center of the face between the IB nodes and the adjacent fluid nodes is modified,
\begin{equation}
f_\nu = \left( \nu  + {\nu_t}^* \right) \left( \frac{\xi_m^i \xi_m^k}{J} \frac{\partial u_j}{\partial \xi^k} + \frac{\xi_k^i \xi_j^m}{J} \frac{\partial u_k}{\partial \xi^m} \right), \quad \nu_t^* = \frac{\left. \nu_t \right|_B + \left. \nu_t \right|_C}{2}.
\label{eq_momentum_flux}
\end{equation}

In this way, the predicted wall shear stress in Eq.~(\ref{eq_FNN_tauw}) is transformed into the momentum flux on the IB boundaries between the IB nodes and fluid nodes. Finally, the modified momentum flux is applied to the governing equation (Eq.~(\ref{eq_1})) to represent the effect of solid boundary on the surrounding flow field.

\subsubsection{Pretrained neural network model for wall shear stress}\label{subsec:tau_w}

The neural network model for estimating wall shear stress is pretrained using the data from the flow over periodic hills and the logarithmic law for the mean streamwise velocity profile~\cite{Zhou_etal_PoF_2023}. The flow over periodic hills, even though its geometry is relatively simple, contains different flow regimes, i.e., flows with pressure gradients, flow separation, and flow reattachment. The inclusion of the logarithmic law in model training then enables the trained model to react properly to flows in the equilibrium state.

The FNN is employed for building the connection between the near-wall flow quantities and the wall shear stress. The neural network comprises an input layer, six hidden layers with 15 neurons in each layer, and an output layer. The hyperbolic tangent (tanh) function serves as the activation function. The input features consist of six flow quantities at three wall-normal points, with the distance between two adjacent points denoted as $\Delta y_n/\delta_0 = 0.03$,
\begin{equation}
X_I = \left\{ \ln\left(\frac{y_n}{y^{*}}\right), \frac{u_{w,t}}{y_n} \frac{\delta_0}{U_b}, \frac{u_{w,n}}{y_n} \frac{\delta_0}{U_b}, \frac{u_s}{y_n} \frac{\delta_0}{U_b}, \frac{\partial p}{\partial w_t} \frac{y_n}{\delta_0} \frac{\delta_0}{U_b^2}, \frac{\partial p}{\partial w_n} \frac{y_n}{\delta_0} \frac{\delta_0}{U_b^2} \right\},
\label{eq_input}
\end{equation}
where $u_{w,t}$, $u_{w,n}$ and $u_s$ are the velocity components in the wall-tangential, wall-normal and spanwise directions, $\frac{\partial p}{\partial w_t}$, $\frac{\partial p}{\partial w_n}$ are the pressure gradients in the wall-tangential and wall-normal directions, $U_b$ is the bulk velocity, $\delta_0$ is the global length scale. It is noted that the length scale $\delta_0$ represents the scale of the outer flow. It is set as the hill height $h$ in the flow over periodic hills, and the half channel width $\delta$ in turbulent channel flows. In this work, it is set as the diameter $D$ of the Suboff body.
The wall-normal distance is normalized by a near-wall length scale $y^{*} = \nu/u_{\tau p}$~\citep{Duprat_etal_PoF_2011}, where $\nu = \mu/\rho$ is the kinematic viscosity, $u_{\tau p} = \sqrt{u_{v}^2+u_p^2}$, $u_{v} = \sqrt{\left| \frac{\nu u_{w, t}}{y_n} \right|}$, $u_p = \left| \frac{\nu}{\rho} \frac{\partial p}{\partial w_t} \right|^{1/3}$.
The output labels are the wall-tangential and spanwise wall shear stresses,
\begin{equation}
Y_O^{(w)} = \left\{ \frac{\tau_{w, t}}{U_b^2}, \frac{\tau_{w, s}}{U_b^2} \right\}.
\label{eq_Tauw_out}
\end{equation}

The cost function is defined as the mean square error between the predicted output and the real output. The error backpropagation (BP) scheme~\citep{Rumelhart_etal_Nature_1986} implemented with TensorFlow~\citep{Google_Brain_2016} is employed to pretrain the FNN model by optimizing the weight and bias coefficients to minimize the cost function. A detailed description of the training procedures can be found in our previous work~\citep{Zhou_He_Yang_PRF_2021}.

In the next section, the neural network for modifying the near-wall momentum flux is trained in the WMLES environment using the ensemble Kalman method.

\section{Embedded training of neural network model based on ensemble Kalman method} \label{sec:ML_strategy}


%
The ensemble Kalman method is a statistical inference method based on Monte Carlo sampling, and has been widely used in various applications~\citep{Chen_etal_NN_2019, zhang2020evaluation, schneider2022ensemble, Liu_Zhang_He_AIAA_2023}. 
In this work, the method is employed to learn the weights in the neural network model for wall shear stress. It utilizes the statistics of the weights and model predictions to compute the gradient and Hessian of the cost function to update the model. The cost function is given as,
\begin{equation}
    J= \| w_m^{n+1} - w_m^n  \|_\mathsf{P}+ \| \mathcal{H}[w_m^{n+1}] - \mathsf{y} \|_\mathsf{R},    
\end{equation}
where $w_m$ is the weight of neural network, $m$ is the sample index, $n$ is the iteration index, $\mathsf{P}$ is the error covariance of neural network model, $\mathsf{R}$ is the error covariance of observation data, $\mathsf{y}$ is the observation data that obey the normal distribution with zero mean and variance of $\mathsf{R}$, and $\mathcal{H}$ is the model operator that maps the neural network weights to model predictions. 

The method uses the ensemble of the neural network samples $W$ ($=[w_1, w_2, \cdots, w_M]$) to estimate the sample mean~$\bar{W}$ and covariance~$\mathsf{P}$ as
\begin{equation}
\left\{
\begin{aligned}
	\bar{W} &= \frac{1}{M} \sum_{m=1}^M w_{m} \text{,} \\
	\mathsf{P} &= \frac{1}{M-1} (W - \bar{W})(W-\bar{W})^\top \text{,}
\end{aligned}
\right.
\end{equation}
where $M$ is the sample size.
Based on the Gauss-Newton method, the first- and second-order derivatives of the cost function are required to update the weights.
The ensemble Kalman method uses the statistics of these samples to estimate the derivative information~\citep{luo2015iterative}.
At the $n^{th}$ iteration, each sample~$w_m$ is updated based on
\begin{equation}
    w_m^{n+1} = w_m^n + \mathsf{PH}^\top (\mathsf{HPH}^\top + \mathsf{R})^{-1} (\mathsf{y}_m^n - \mathsf{H}w_m^n) \text{,}
    \label{eq:enkf}
\end{equation}
where $\mathsf{H}$ is the tangent linear model operator.
The readers are referred to Ref.~\cite{Zhang_etal_JFM_2022} for details of the employed ensemble Kalman method.

\begin{figure}[!ht]
    \centering
    \includegraphics[width=1.0\textwidth]{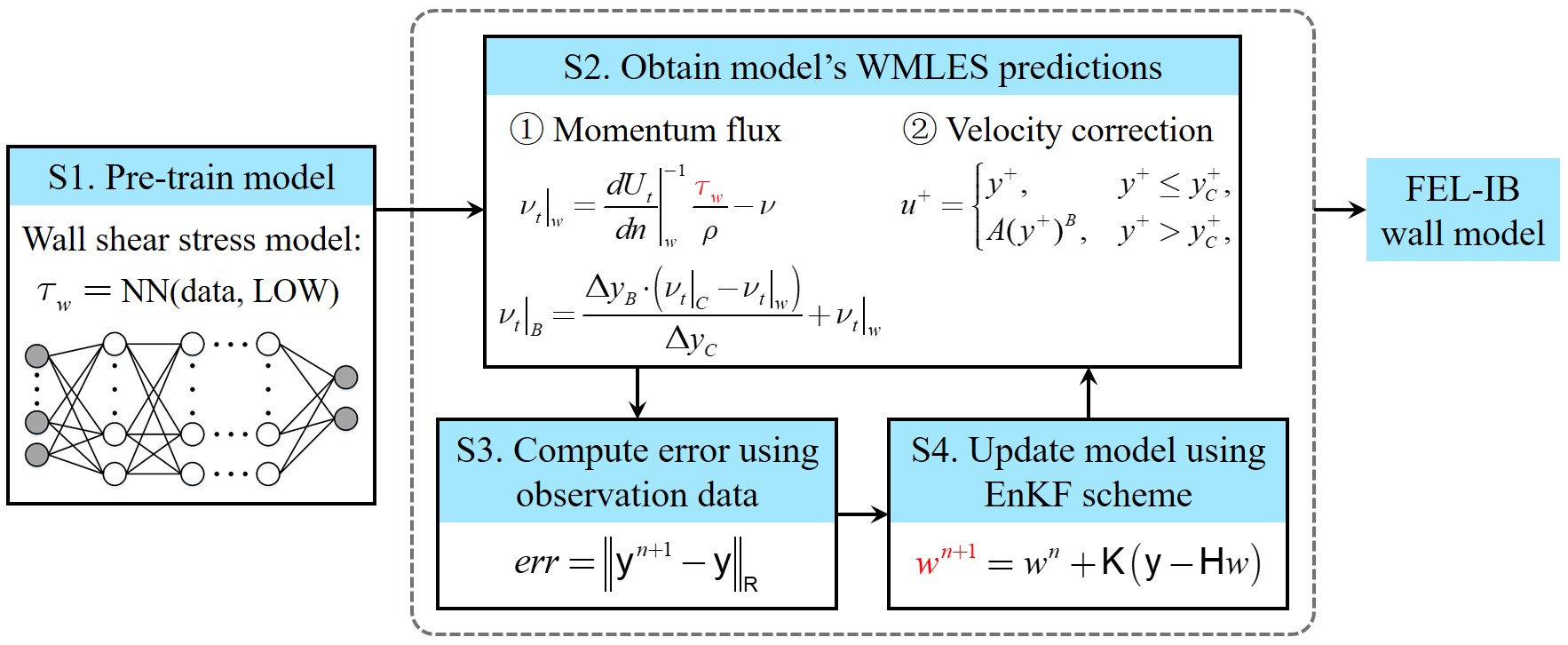}
    \caption{Embedded training of the FEL-IB wall model, where EnKF denotes the ensemble Kalman filter.}
    \label{fig:schematic}
\end{figure}

The procedure for embedded training of the neural network-based wall model is divided into the following five steps:
\setlist[enumerate,1]{label=(\arabic*),font=\textup}
\begin{enumerate}
    \item Initial sampling of the neural network model. The neural network weight $w_m$ for each sample are determined by sampling from a Gaussian distribution with a mean of $w_0$ and a user-specified standard deviation. The pretrained wall shear stress model in section~\ref{subsec:tau_w} is employed as the initial weight $w_0$.
    \item Obtain the WMLES predictions for an ensemble of neural network models. This step is done by applying the near-wall velocity reconstruction model and the modified model for near-wall momentum flux based on neural network to the WMLES cases.
    \item Compute the error of the model predictions. The WMLES predictions and the observation data are employed to compute the errors. 
    \item Update the neural networks using the ensemble Kalman method. The errors obtained in the last step are employed to update the weights of the neural network model for wall shear stress using Eq.~(\ref{eq:enkf}).
    \item Iterate steps (2) to (4) until the maximum number of iterations is reached.
\end{enumerate}

\section{Test cases}\label{sec:Test_cases}

\subsection{Flow over a body of revolution}\label{sec:4.3}
\subsubsection{Case setup}
\begin{figure}[!ht]
\centering{\includegraphics[width=0.9\textwidth]{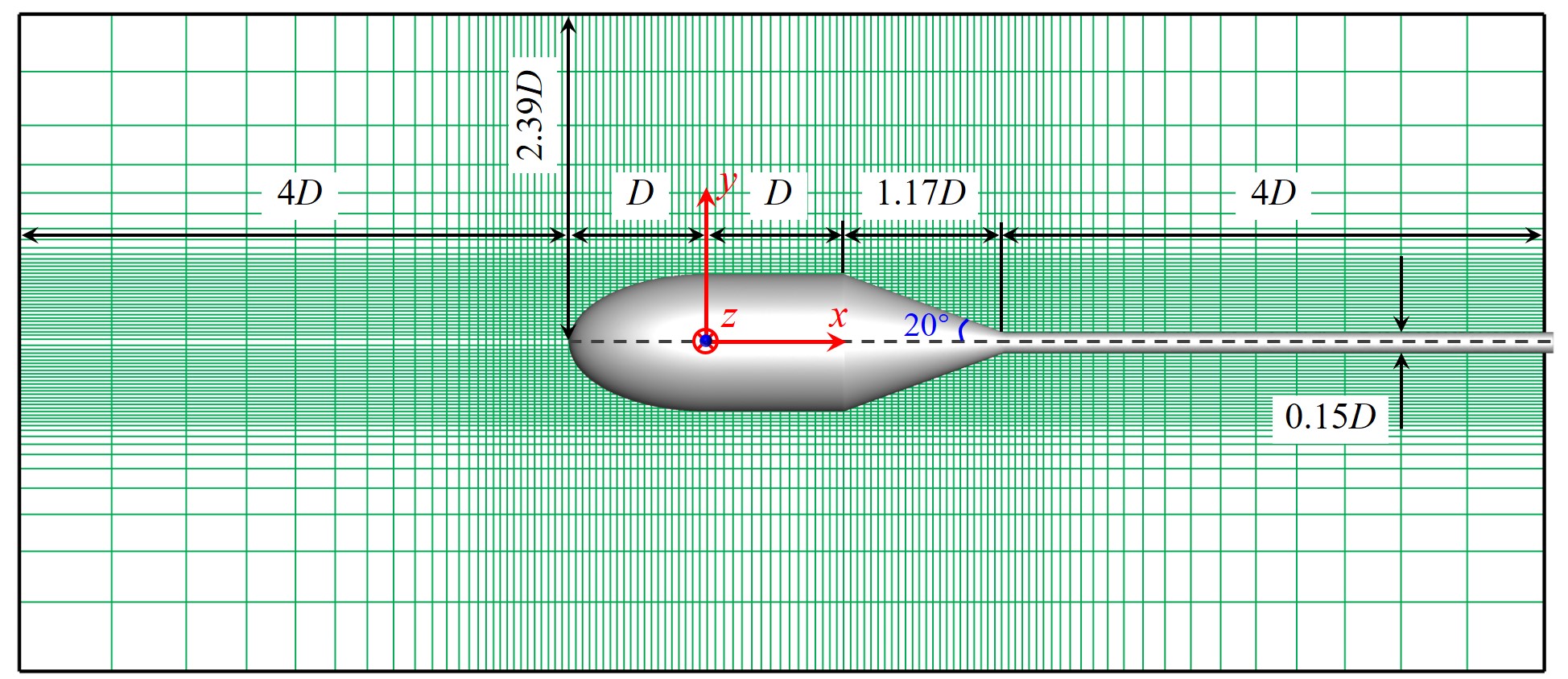}}
  \caption{Schematic of the geometry for the body of revolution and computational domain with the $xOy$ plane of the Cartesian background grid (every fifth nodes).}
\label{fig:grid_BOR}
\end{figure}
The proposed model is first applied to the flow over a body of revolution (BOR)~\cite{Zhou_etal_AIAA_2020}, as shown in figure~\ref{fig:grid_BOR}. The BOR
consists of a cylindrical section in the middle, a 2:1 ellipsoidal nose, and a tail cone that connects to the centerbody at one end and a cylindrical support pole at the other end. The diameter $D$ of the centerbody is used as the length scale for normalization. The nose and the centerbody both have a length equal to $D$, the tail cone has a $20^{\circ}$ half apex angle and a length of $1.17D$, and the support pole has a diameter of $0.15D$. The simulation is conducted in a cuboid computational domain with $11.17D\times 4.78D \times 4.78D$ at the streamwise, vertical and spanwise directions, discritized with a grid resolution of $665\times 370 \times 370$. The Reynolds number is defined as $Re_L = U_\infty L/\nu = 1.9\times 10^6$, where $U_\infty$ is the free-stream velocity, $L=3.17D$ is the total length of the BOR. The BOR is placed at zero angle of attack with respect to the inflow. At the inlet, a uniform inflow is imposed. The Neumann boundary condition is employed at the outlet. As for the computational boundaries in the spanwise and vertical directions, the free-slip boundary condition is imposed.
In the simulation, a circumferential trip ring is positioned close to the downstream end of the nose ($x_t/D=0.0$) in order to induce transition in the boundary layer.

\subsubsection{Results}

\begin{figure}[!ht]
\centering{\includegraphics[width=0.48\textwidth]{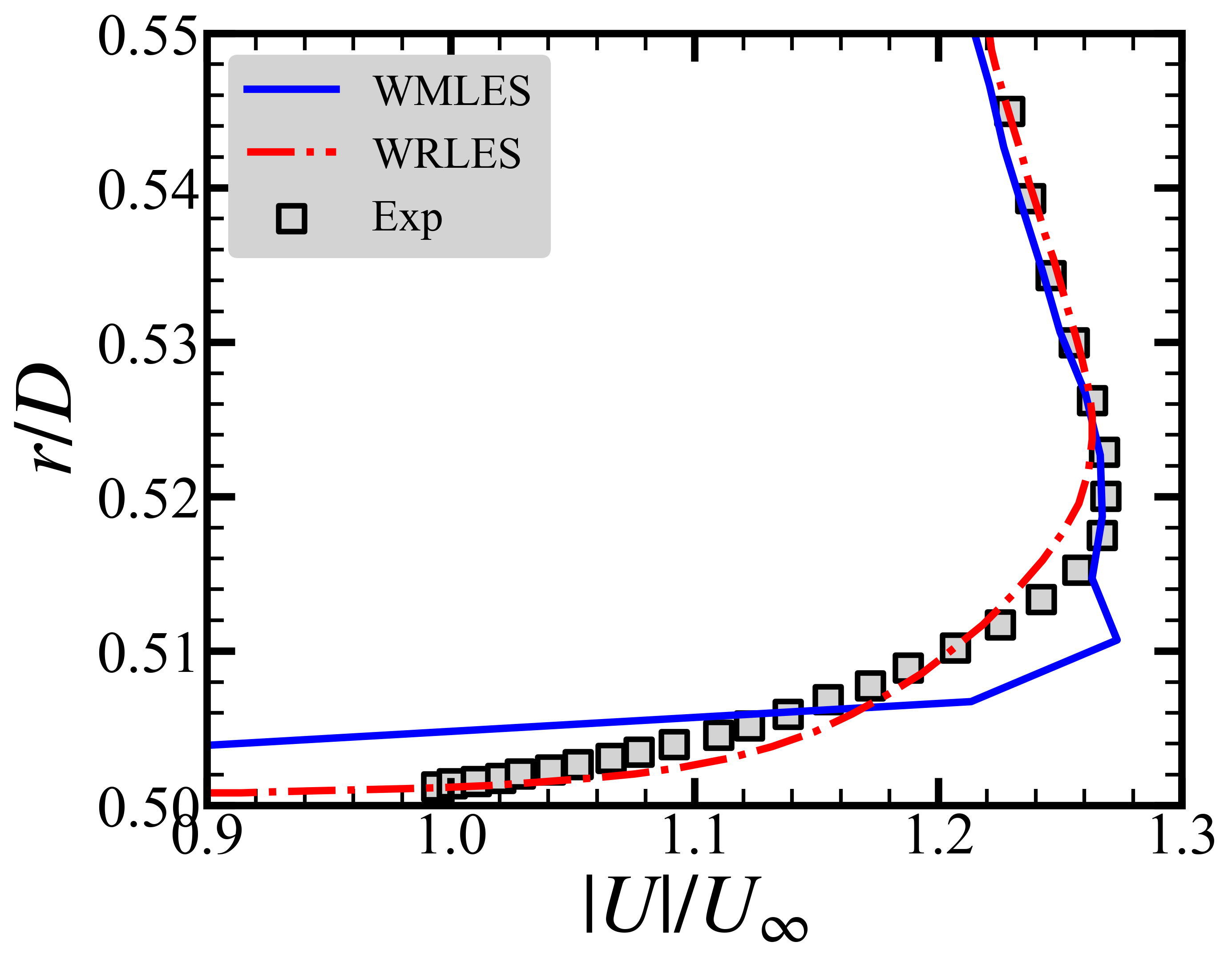}}
  \caption{Comparison of the magnitude of time-averaged velocity at the location $x/D=0.98$ for the BOR with $Re_L=1.9\times 10^6$ obtained from the present WMLES with the FEL-IB model, the WRLES of Zhou \textit{et al.}~\cite{Zhou_etal_AIAA_2020} and the experiment of Hickling \textit{et al.}~\cite{Hickling_etal_AIAA_2019}.}
\label{fig:BOR_velo_1}
\end{figure}
\begin{figure}[!ht]
\centering
\begin{subfigure}[b]{0.328\textwidth}
	\centering
	\includegraphics[width = 1.0\textwidth]{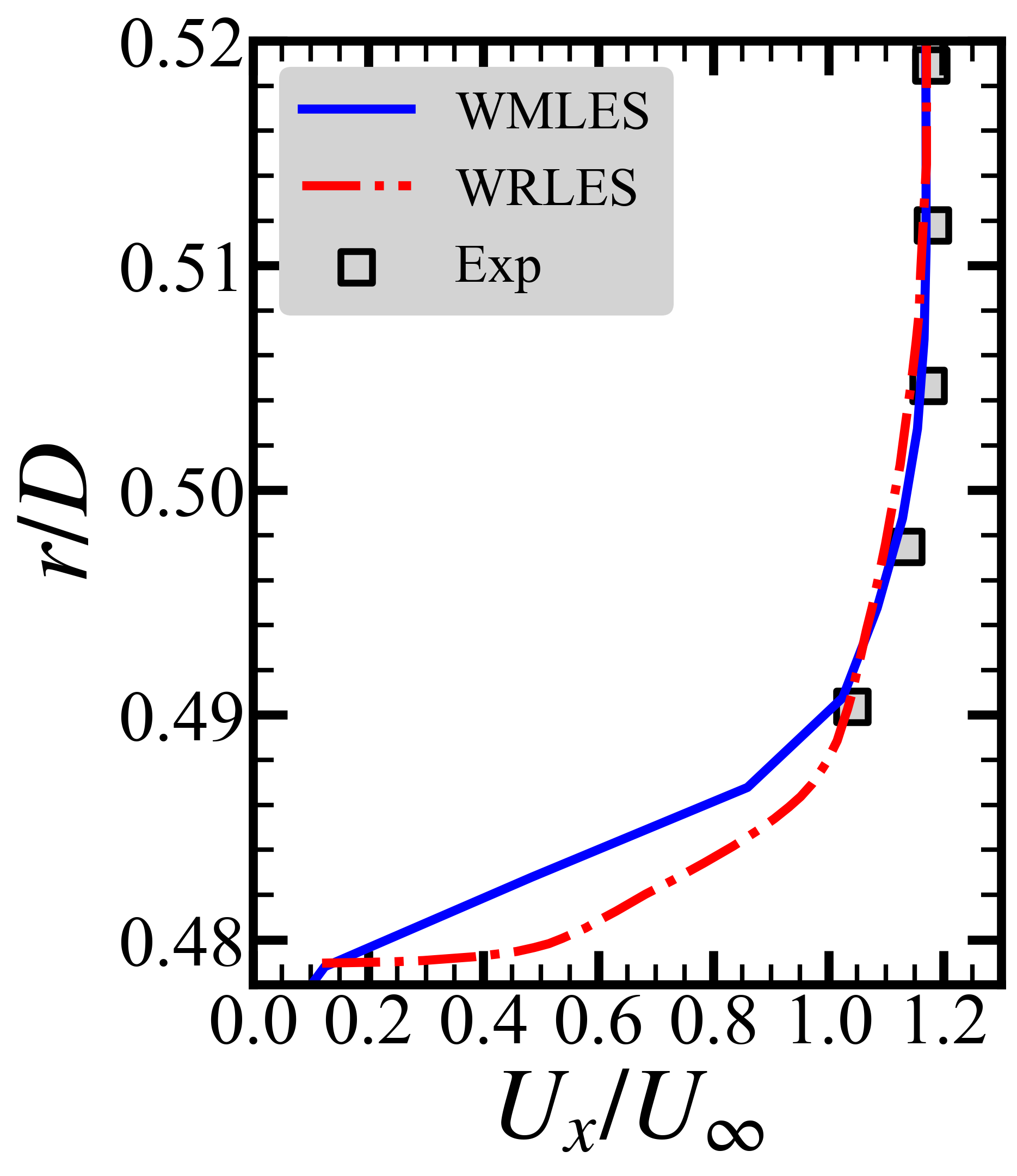}
	\subcaption{$x/D$ = 1.06}
\end{subfigure}
\begin{subfigure}[b]{0.328\textwidth}
	\centering
	\includegraphics[width = 1.0\textwidth]{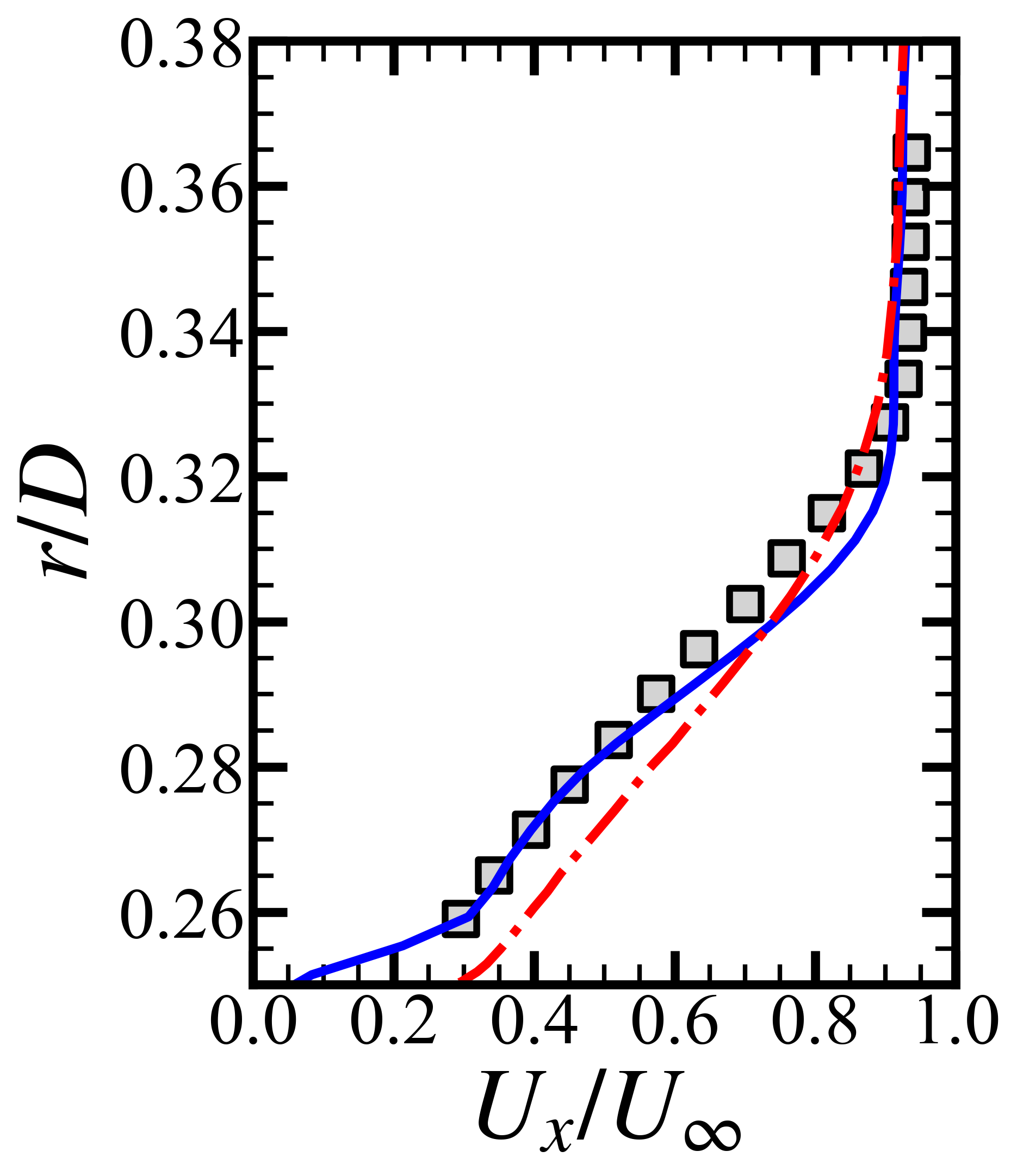}
	\subcaption{$x/D$ = 1.69}
\end{subfigure}
\begin{subfigure}[b]{0.328\textwidth}
	\centering
	\includegraphics[width = 1.0\textwidth]{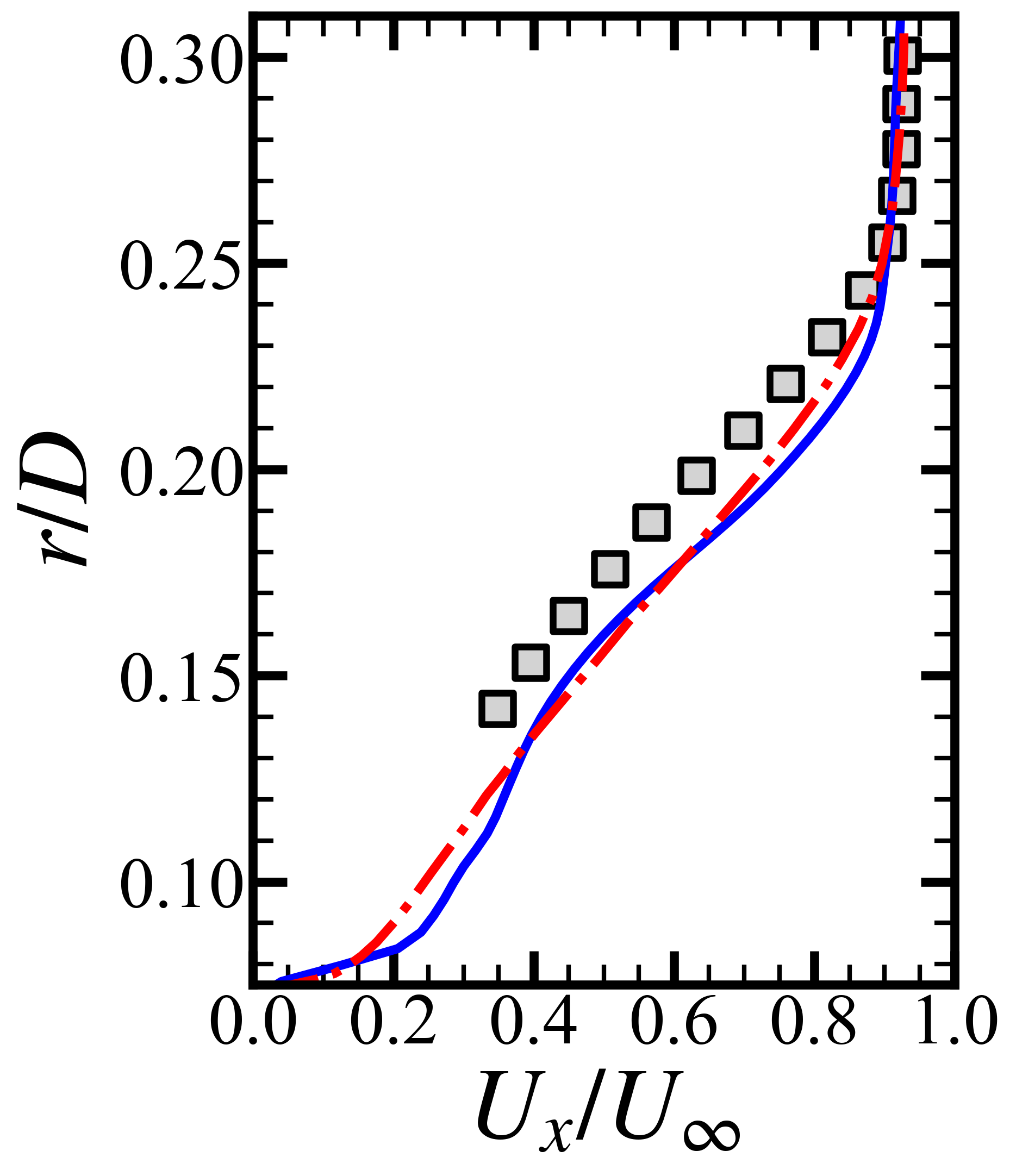}
	\subcaption{$x/D$ = 2.17}
\end{subfigure}
\caption{Comparison of the profiles of time-averaged axial velocity for the BOR at $Re_L=1.9\times 10^6$ obtained from the present WMLES with the FEL-IB model, the WRLES of Zhou \textit{et al.}~\cite{Zhou_etal_AIAA_2020} and the experiment of Hickling \textit{et al.}~\cite{Hickling_etal_AIAA_2019} at three axial locations $x/D=1.06$, 1.69 and 2.17.}
\label{fig:BOR_velo_2}
\end{figure}

The boundary layer profile at $0.02D$ upstream of the tail-cone ($x/D=0.98$) is illustrated in figure~\ref{fig:BOR_velo_1} in terms of the magnitude of time-averaged velocity. Compared with the data from the WRLES of Zhou \textit{et al.}~\cite{Zhou_etal_AIAA_2020} and the experiment of Hickling \textit{et al.}~\cite{Hickling_etal_AIAA_2019}, the WMLES with the FEL-IB model basically predict the variation tendency of the velocity profile, but the thickness of boundary layer is slightly underestimated. At further downstream flow, the vertical profiles of time-averaged axial velocity at three axial stations ($x/D=1.06$, 1.69 and 2.17) on the tail cone are shown in figure~\ref{fig:BOR_velo_2}. The WMLES results with the FEL-IB model are in good agreement with those from the WRLES and experimental results.

\subsection{Flow over the DARPA Suboff model}\label{sec:4.4}
\subsubsection{Case setup}

In this section, the proposed model is systematacially examined in the flow over the DARPA Suboff model.
As seen in Figure~\ref{fig:geometry}, the bare hull of the DARPA Suboff is axis-symmetric and consists of a streamlined forebody, a parallel middle body, and a stern with contraction in the radial direction. The Reynolds number based on the length of the Suboff $L = 8.6D$ is equal to $Re=U_0 L/\nu=1.2 \times 10^6$, where $U_0$ denotes the velocity of incoming flow, $\nu$ denotes the kinematic viscosity of the fluid, $D$ denotes the diameter of the middle body of the Suboff. At the inlet, a uniform inflow is imposed. Free-slip boundary condition is imposed on the computational boundaries in the spanwise and vertical directions. At the outlet, the Neumann boundary condition is employed.
\begin{figure}[!ht]
\centering{\includegraphics[width=1.0\textwidth]{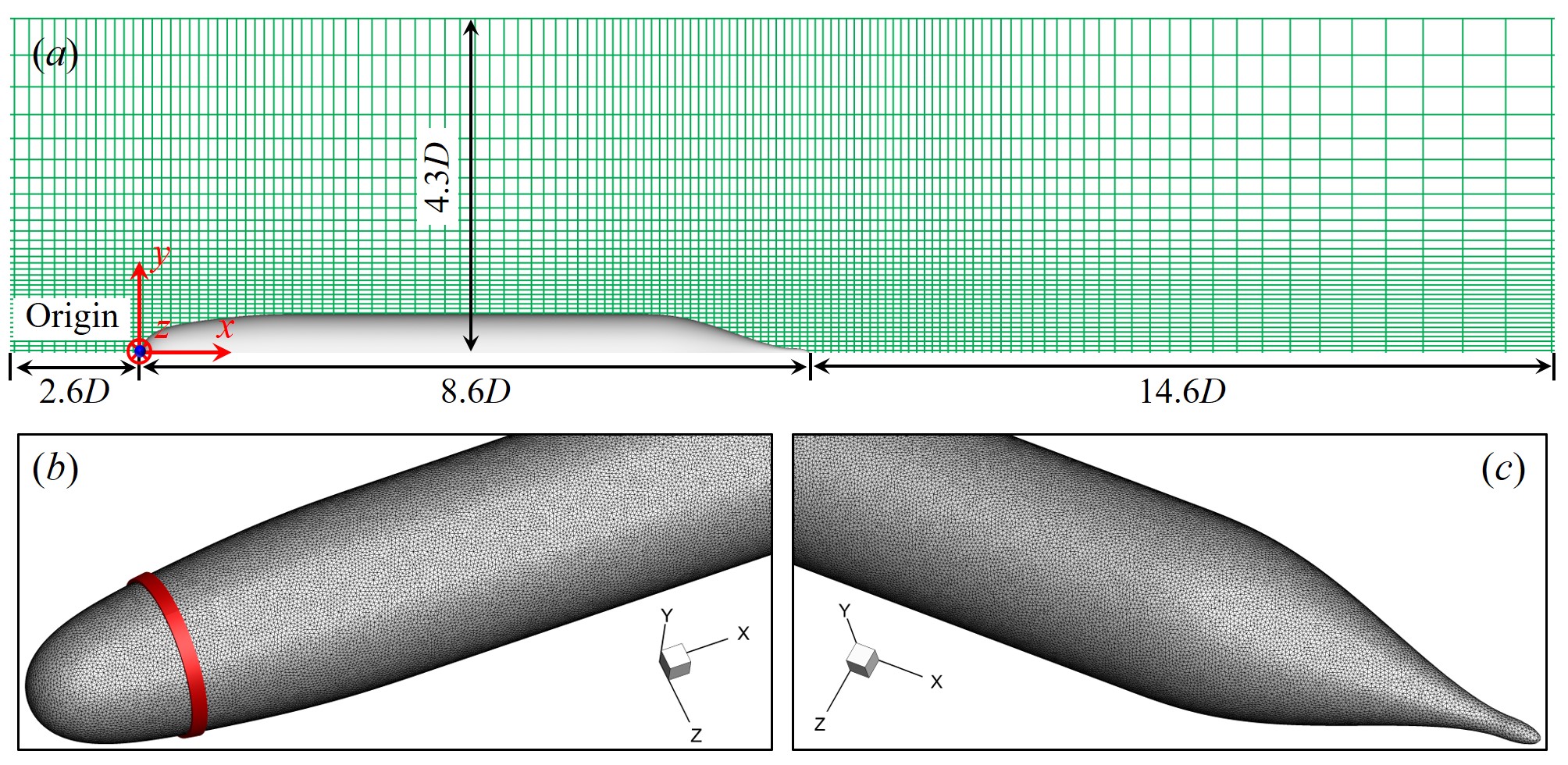}}
  \caption{(a) Computational domain with the $xOy$ plane of the Cartesian background grid (1 in every 5 nodes is shown) and the Suboff model; (b) the unstructured surface mesh for the streamlined forebody with a trip wire at the location $x_t/D=0.75$ of the hull surface; (c) the unstructured surface mesh for the stern of Suboff model.}
\label{fig:grid_Suboff}
\end{figure}
The Suboff model is placed at zero angle of attack and zero yaw angle with respect to the inflow, as shown in Figure~\ref{fig:grid_Suboff}. The origin of the coordinate system coincides with the nose of the hull. In this coordinate system, the extent of the computational domain in the streamwise, vertical and spanwise directions is [$-2.6D$, $23.2D$]$\times$[$-4.3D$, $4.3D$]$\times$[$-4.3D$, $4.3D$]. In this work, we carried out two simulated cases with different grids, for which the space intervals and number of grids are in different regions are shown in Tables~\ref{tab:Grid_case}-\ref{tab:G2_setup}. Figure~\ref{fig:grid_Suboff} shows the distribution of G1 grid in table~\ref{tab:G1_setup}, which is locally refined near the Suboff model and stretched outwards.
\newcommand{\tabincell}[2]{\begin{tabular}{@{}#1@{}}#2\end{tabular}}
\begin{table}[!ht]
\centering
\caption{\label{tab:Grid_case}.The grid resolution for different simulated cases.}
\begin{tabular}{p{1.6cm}<{\centering}p{2.5cm}<{\centering}p{2.5cm}<{\centering}}
  \hline
  Case & $N_x\times N_y\times N_z$    &  Grid numbers     \\
  \hline
  G1  &  $275 \times 130 \times 130$  & $4.65\times 10^6$ \\
  G2  &  $600 \times 400 \times 400$  & $9.60\times 10^7$ \\
  \hline
\end{tabular}
\end{table}
\begin{table}[!ht]
\centering
\caption{\label{tab:G1_setup} Details of the space interval ($\Delta h$) and number of grids ($N$) at different locations for the G1 grid ($275 \times 130 \times 130$). For the space interval, ``u'' denotes the uniform grid, ``r'' and ``l'' denote the non-uniform grid defined using the tanh function, with the smallest grid cell on the right and left side, respectively. }
\begin{tabular}{c|c|c|c|c|c}
  \hline\hline
  $x/D \in$ & [-2.6, -0.4] &  [-0.4, 2.0] &  [2.0, 6.0] & [6.0, 10.0] & [10.0, 23.2] \\
  \hline
  $N$  &  25         &  40      &  50        & 80       & 80 \\
  \hline
  $\Delta h/D$   &  0.06 (r)    &  0.06 (u)    &  \makecell{0.06 (l) \\ 0.05 (r)}   & 0.05 (u)    & 0.05 (l)   \\
  \hline\hline
  $y/D \in$ & [-4.3, -0.75] &  \multicolumn{3}{c|}{[-0.75, 0.75]}  & [0.75, 4.3] \\
  \hline
  $N$  &  40         &  \multicolumn{3}{c|}{50}       & 40  \\
  \hline
  $\Delta h/D$   &  0.03 (r)   &  \multicolumn{3}{c|}{0.03}   & 0.03 (l) \\
  \hline\hline
  $z/D \in$ & [-4.3, -0.75] &  \multicolumn{3}{c|}{[-0.75, 0.75]}  & [0.75, 4.3]  \\
  \hline
  $N$  &  40         &  \multicolumn{3}{c|}{50}  & 40 \\
  \hline
  $\Delta h/D$   &  0.03 (r)   &  \multicolumn{3}{c|}{0.03}  & 0.03 (l) \\
  \hline\hline
\end{tabular}
\end{table}
\begin{table}[!ht]
\centering
\caption{\label{tab:G2_setup} Details of the space interval ($\Delta h$) and number of grids ($N$) at different locations for the G2 grid ($600 \times 400 \times 400$).}
\begin{tabular}{c|c|c|c|c|c}
  \hline\hline
  $x/D \in$ & [-2.6, -0.4] &  [-0.4, 2.0] &  [2.0, 6.4] & [6.4, 10.0] & [10.0, 23.2] \\
  \hline
  $N$  &  50         &  80      &  120        & 180       & 170 \\
  \hline
  $\Delta h/D$   &  0.03 (r)    &  0.03 (u)    &  \makecell{0.03 (l) \\ 0.02 (r)}   & 0.02 (u)    & 0.02 (l)   \\
  \hline\hline
  $y/D \in$ & [-4.3, -0.75] &  \multicolumn{3}{c|}{[-0.75, 0.75]}  & [0.75, 4.3] \\
  \hline
  $N$  &  120         &  \multicolumn{3}{c|}{160}       & 120  \\
  \hline
  $\Delta h/D$   &  0.005 (r)   &  \multicolumn{3}{c|}{0.005}   & 0.005 (l) \\
  \hline\hline
  $z/D \in$ & [-4.3, -0.75] &  \multicolumn{3}{c|}{[-0.75, 0.75]}  & [0.75, 4.3]  \\
  \hline
  $N$  &  120         &  \multicolumn{3}{c|}{160}  & 120 \\
  \hline
  $\Delta h/D$   &  0.005 (r)   &  \multicolumn{3}{c|}{0.005}  & 0.005 (l) \\
  \hline\hline
\end{tabular}
\end{table}


To promote the generation of turbulent boundary layer, a wall-normal velocity perturbation is imposed at the location $x_t/D=0.75$ of the hull surface to mimic a trip wire, as shown in figure~\ref{fig:grid_Suboff}(b). The functional form of the velocity perturbation is $u_t = 0.06U_0 + 0.02U_0\cdot \alpha$, where $\alpha$ is a random number at the range of $[-1, 1]$.



\subsubsection{Evaluation of the FEL-IB model for the Suboff case at a low Reynolds number}
\begin{figure}[!ht]
\begin{minipage}[t]{1.0\linewidth}
    \centering
    \includegraphics[width=0.9\textwidth]{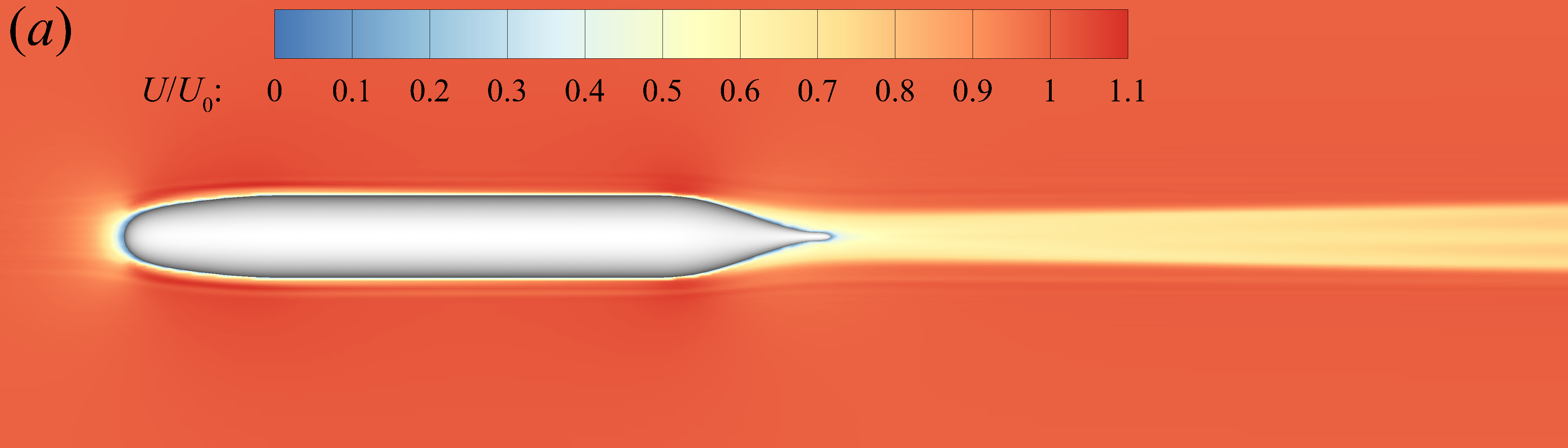}
    \label{fig:slice_G1_velo}
    \vspace{1mm}
\end{minipage}
\begin{minipage}[t]{1.0\linewidth}
    \centering
    \includegraphics[width=0.9\textwidth]{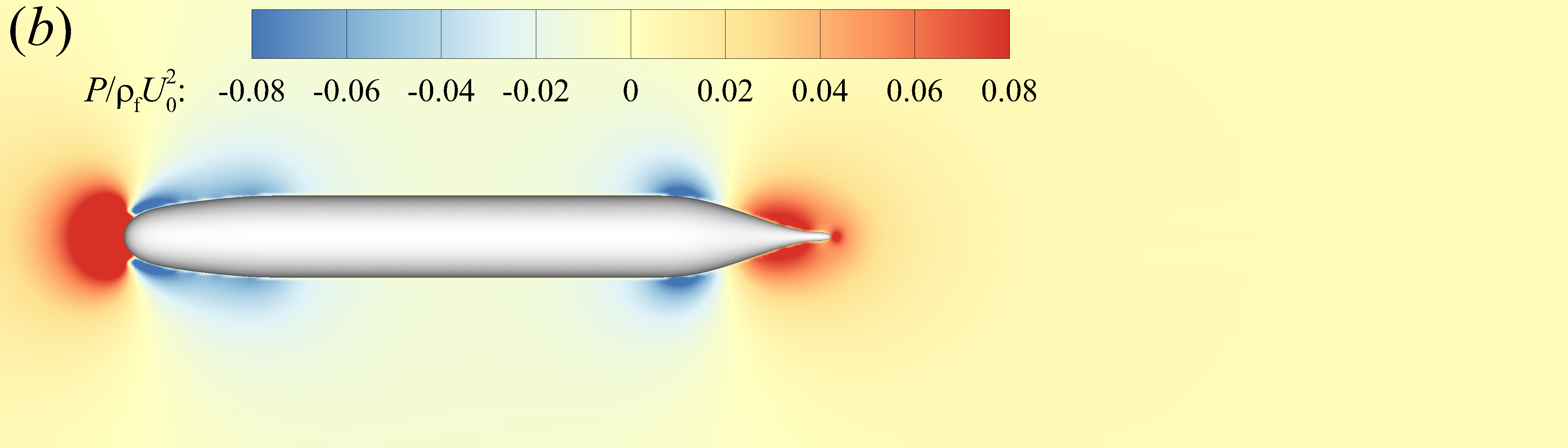}
    \label{fig:slice_G1_pres}
    \vspace{1mm}
\end{minipage}
  \caption{(a) Time-averaged streamwise velocity and (b) time-averaged pressure on the 2D slice $z=0.0$ of the flow over the Suboff with G1 grid and $Re_L=1.2\times 10^6$. }
\label{fig:slice_G1}
\end{figure}
\begin{figure}[!ht]
\centering
\begin{subfigure}[b]{0.48\textwidth}
	\centering
	\includegraphics[width = 1.0\textwidth]{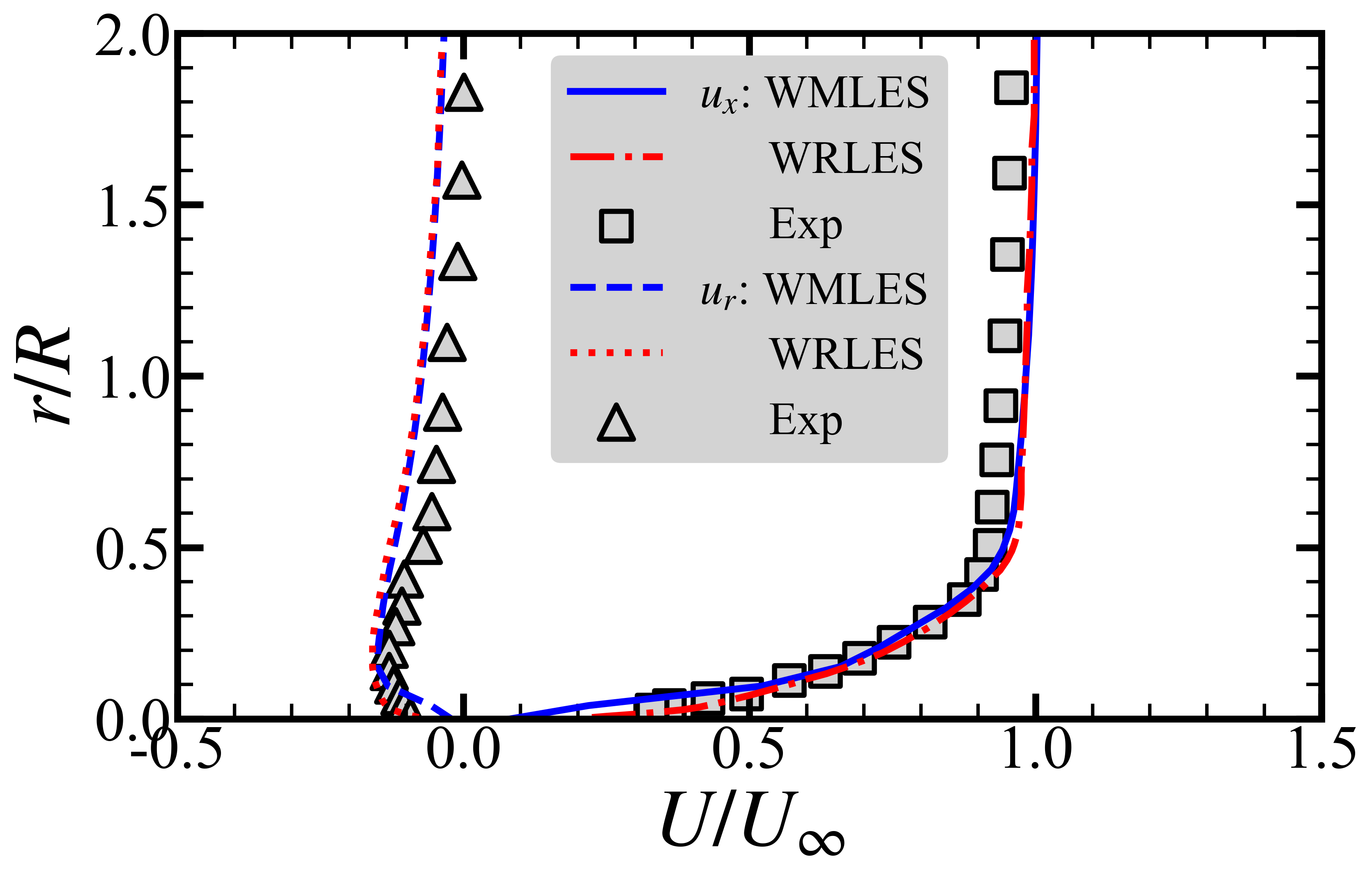}
	\subcaption{$x/L$ = 0.904}
\end{subfigure}
\begin{subfigure}[b]{0.48\textwidth}
	\centering
	\includegraphics[width = 1.0\textwidth]{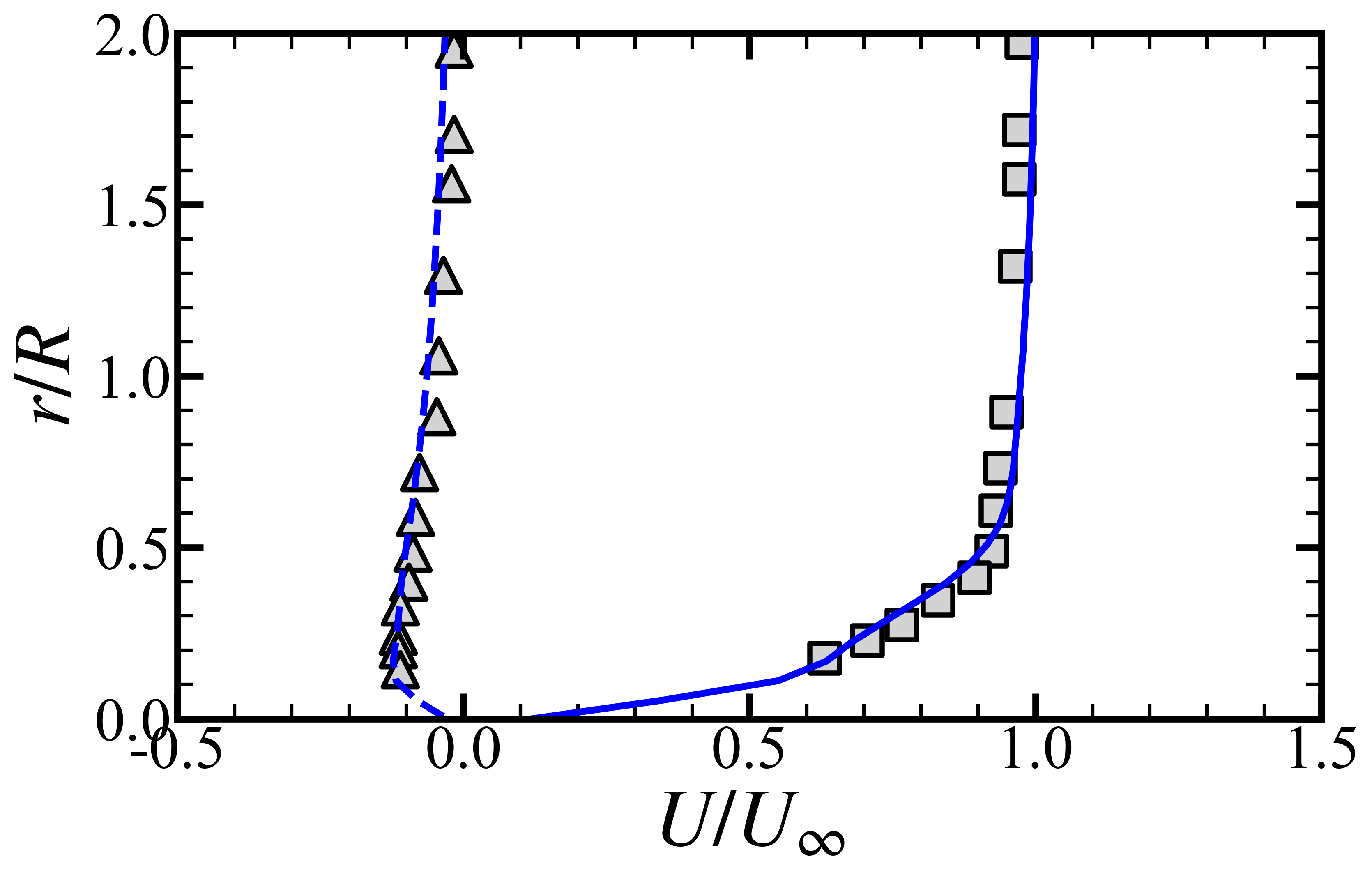}
	\subcaption{$x/L$ = 0.927}
\end{subfigure}
\begin{subfigure}[b]{0.48\textwidth}
	\centering
	\includegraphics[width = 1.0\textwidth]{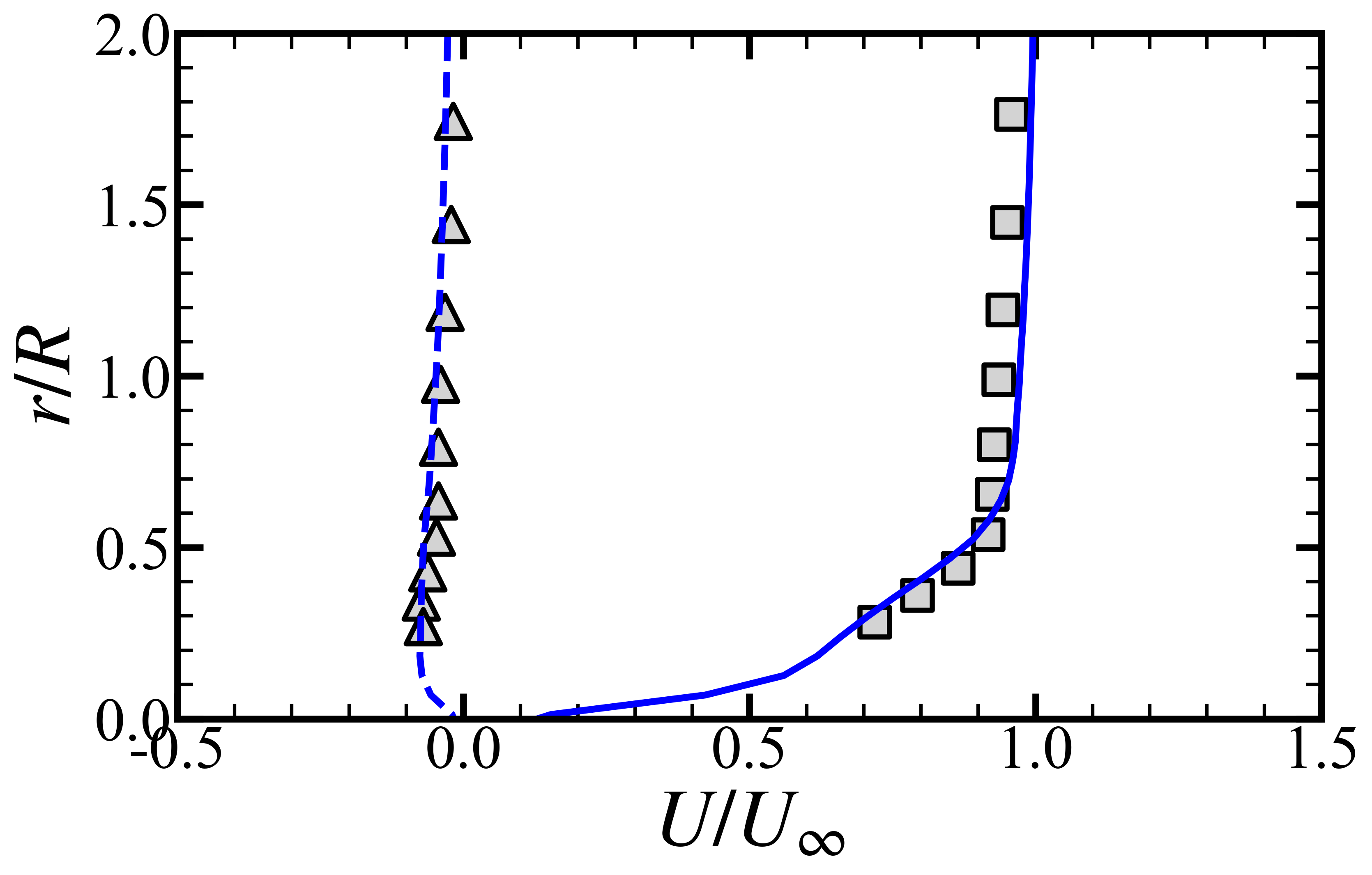}
	\subcaption{$x/L$ = 0.956}
\end{subfigure}
\begin{subfigure}[b]{0.48\textwidth}
	\centering
	\includegraphics[width = 1.0\textwidth]{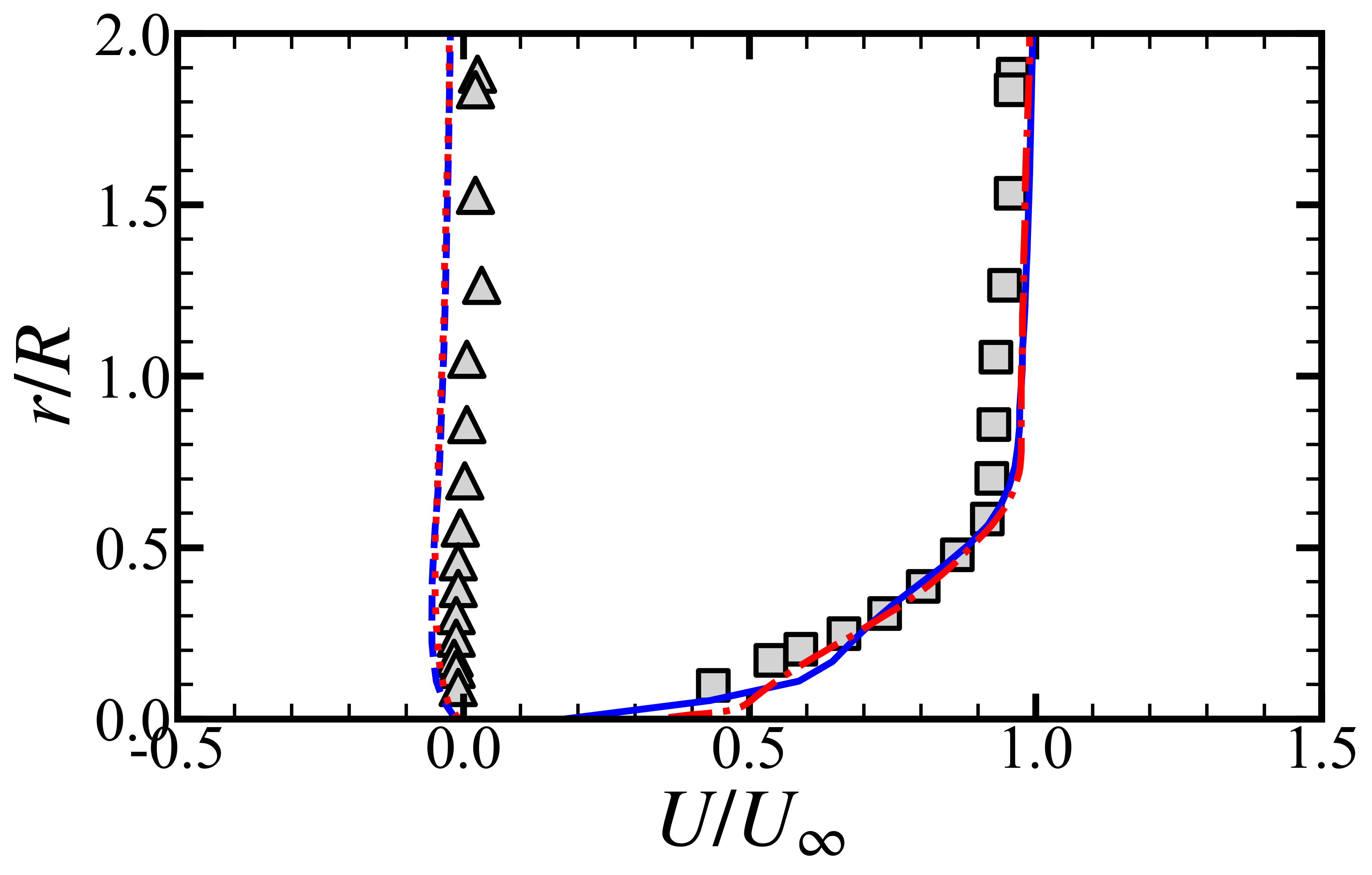}
	\subcaption{$x/L$ = 0.978}
\end{subfigure}
\caption{Comparison of time-averaged axial and radial velocity profiles for the Suboff case at $Re_L=1.2\times10^6$ obtained from the present WMLES with G1 grid, the WRLES of Kumar \& Mahesh~\cite{Kumar_Mahesh_JFM_2018} and the experiment of Huang \textit{et al.}~\cite{Huang_1994_measurements}.}
\label{fig:velo_avg_G1}
\end{figure}

To evaluate the proposed FEL-IB model, the WMLES of flow over the Suboff at $Re_L=1.2\times 10^6$ is first carried out using the G1 coarse grid. Figure~\ref{fig:slice_G1} plots the contours of the time-averaged streamwise velocity $U$ and pressure $P$ at the two-dimensional (2D) slice $z=0.0$. It is observed that the time-averaged streamwise velocity field around the Suboff is quite simple, while the boundary layer is thin at the middle body and becomes thicker at the curved stern with contraction. The wake flow with velocity deficit downstream the Suboff model could spread to a far field. As for the time-averaged pressure, the positive peaks are observed near $x/D=0.0$ of the forebody and $x/D=8.0$ of the stern, while the negative peaks are observed near the border region between the curved surface and flat surface. Thus the favorable and adverse pressure gradient appears to the forebody and stern of the Suboff, respectively.

Figure~\ref{fig:velo_avg_G1} compares the time-averaged axial and radial velocity profiles from the present WMLES with G1 grid, the WRLES of Kumar \& Mahesh~\cite{Kumar_Mahesh_JFM_2018} and the experiment of Huang \textit{et al.}~\cite{Huang_1994_measurements}. In the figure, $r$ denotes the wall-normal distance, and $r/R=0.0$ represents the surface of the Suboff. The four profiles are all located at the curved stern of the Suboff. We can easily observe that the axial velocity profiles from the WMLES are consistent with those from the WRLES and experiment. The radial velocity profiles from the WMLES exhibit some deviation with the experimental results at $x/L$ = 0.904 and 0.978, but coincide well with the WRLES results.

\begin{table}[!ht]
\centering
\caption{\label{tab:Err_U_G1}The relative errors of time-averaged axial velocity between the WMLES with G1 grid and the WRLES/experiment for the Suboff case at $Re_L=1.2\times10^6$.}
\begin{tabular}{p{1.6cm}<{\centering}p{2.5cm}<{\centering}p{2.5cm}<{\centering}}
  \hline
  $x/L$ & $Err_U$-exp & $Err_U$-WRLES  \\
  \hline
  0.904  &  4.11\%  &  1.51\%  \\
  0.927  &  2.30\%  &  --      \\
  0.956  &  3.42\%  &  --      \\
  0.978  &  4.06\%  &  0.86\%  \\
  \hline
\end{tabular}
\end{table}

%
%

To quantify the prediction accuracy of the proposed model, we introduce the relative error for the flow statistics as follows,
\begin{equation}
Err_f = \frac{\sum |f_{\text{WM}}-f_{\text{ref}}| \Delta y}{\sum |f_{\text{ref}}| \Delta y},
\label{eq_err}
\end{equation}
where $f_{\text{WM}}$ and $f_{\text{ref}}$ represent the flow statistics from the WMLES and reference case (which is WRLES or experiment for the Suboff case), respectively, $\sum$ denotes the integral along the vertical direction.
Table~\ref{tab:Err_U_G1} lists the relative errors of time-averaged axial velocity ($Err_U$) between the WMLES with G1 grid and the WRLES/experiment, which are calculated by Eq.~\ref{eq_err}. The present WMLES shows an error smaller than 5\% at the four streamwise locations compared to the experiment, and another error smaller than 2\% at $x/L=0.904$ and 0.978 compared to the WRLES.

\begin{figure}[!ht]
\centering
\begin{subfigure}[b]{0.48\textwidth}
	\centering
	\includegraphics[width = 1.0\textwidth]{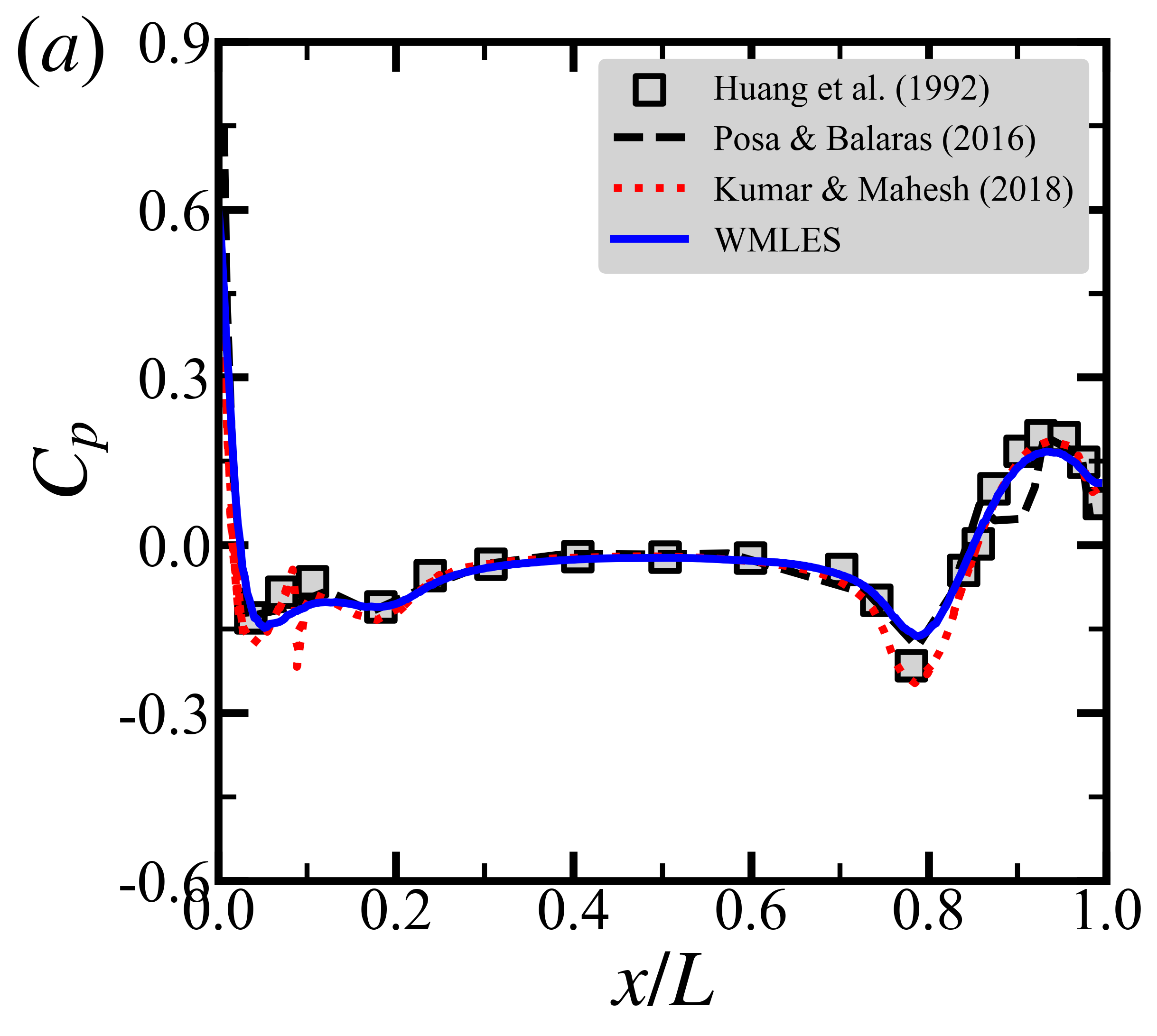}
\end{subfigure}
\begin{subfigure}[b]{0.48\textwidth}
	\centering
	\includegraphics[width = 1.0\textwidth]{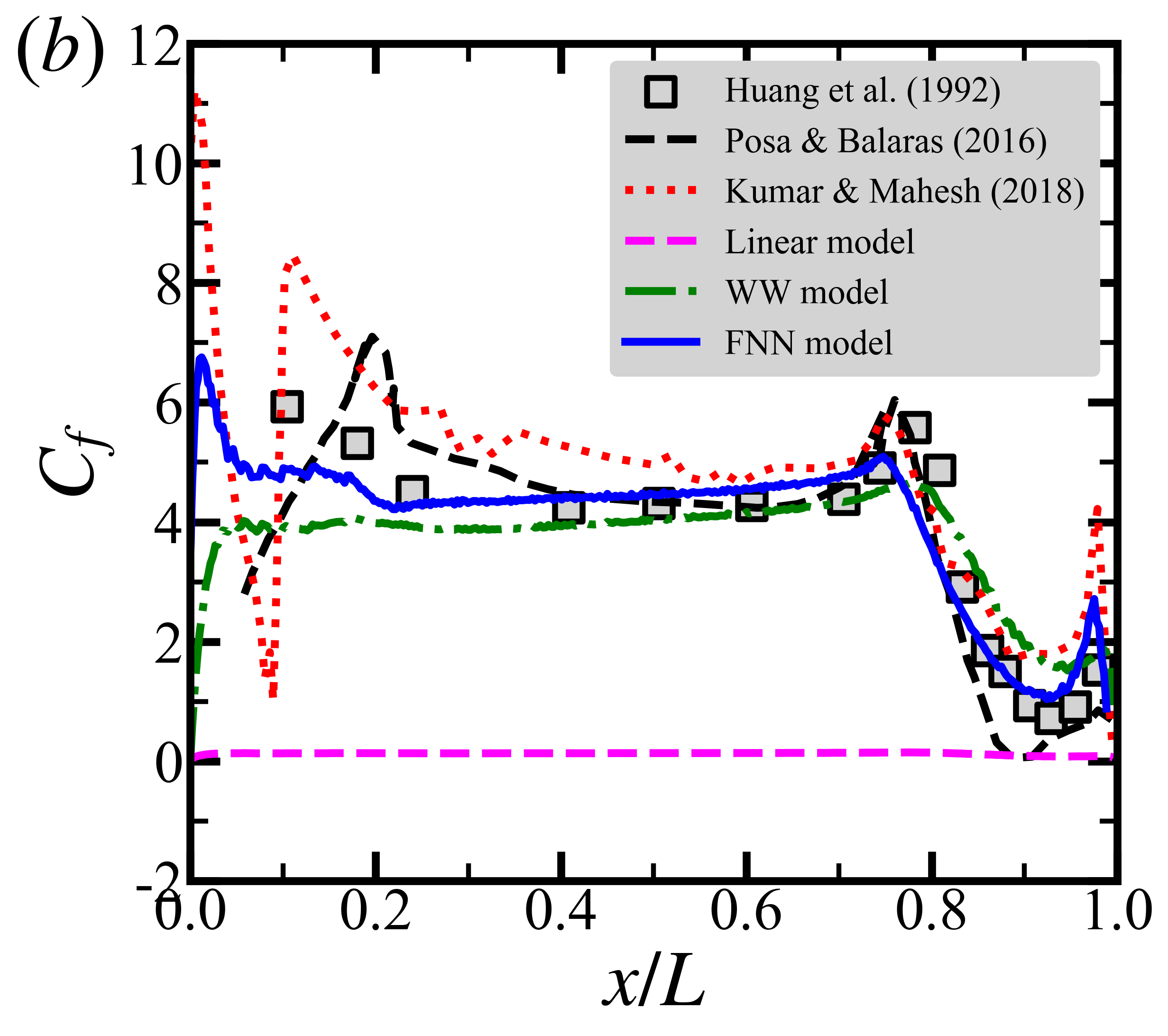}
\end{subfigure}
\caption{Comparison of pressure coefficient $C_p$ and skin-friction coefficient $C_f$ for the Suboff case at $Re_L=1.2\times10^6$ obtained from the present WMLES with G1 grid, the WRLES of Posa \& Balaras~\cite{Posa_Balaras_JFM_2016} and Kumar \& Mahesh~\cite{Kumar_Mahesh_JFM_2018}, and the experiment of Huang \textit{et al.}~\cite{Huang_1994_measurements}.}
\label{fig:Cp_Cf_G1}
\end{figure}

Furthermore, we examine the performance of the FEL-IB model on predicting the skin-friction coefficient $C_f$ and pressure coefficient $C_p$, which are calculated based on the time-averaged velocity and pressure fields. The two coefficients are defined as follows:
\begin{equation}
C_p = \frac{P-P_{\text{ref}}}{\frac{1}{2}\rho U_0^2}, \quad\quad C_f = \frac{\left\langle \tau_w \right\rangle}{\frac{1}{2}\rho U_0^2},
\label{eq_Cf_Cp}
\end{equation}
where $P_{\text{ref}}=0$ is the reference pressure in the far field, $\tau_w$ is the wall shear stress. Figure~\ref{fig:Cp_Cf_G1} compares the pressure and skin-friction coefficients obtained from the present WMLES with G1 grid, two WRLES cases~\cite{Posa_Balaras_JFM_2016, Kumar_Mahesh_JFM_2018}, and the experiment of Huang \textit{et al.}~\cite{Huang_1994_measurements}. Here, the WRLES data of Posa \& Balaras~\cite{Posa_Balaras_JFM_2016} is obtained from the 2D slice $y+z=0.0$ of the Suboff model with appendages. In figure~\ref{fig:Cp_Cf_G1}(a), the pressure coefficient from the WMLES basically agrees with the different referenced results. For the skin-friction coefficient in figure~\ref{fig:Cp_Cf_G1}(b), the two WRLES curves exhibit significant deviations with the experimental curve, especially at the stern of the Suboff. The present WMLES case with G1 grid provides three different predictions of $C_f$. The linear model, which represents the linear interpolation of the computational grids, significantly underestimates the value of $C_f$ at all streamwise locations. The WW model accurately predicts the value of $C_f$ at the parallel middle body of the Suboff, but exhibits some overestimation at the stern. The FNN model pretrained in our previous work~\cite{Zhou_etal_PoF_2023} provides a more closer prediction of $C_f$ to the experimental result than the WW model and the WRLES cases, especially at the stern with adverse pressure gradient.

In summary, the present WMLES with the proposed model accurately predicts the time-averaged velocity field even when the grid number is two orders of magnitude smaller than the WRLES. And the FNN model trained using the data from the periodic hill case and logarithmic law could be generalized to predict the wall shear stress of the Suboff case.

\subsubsection{Evaluation of the FEL-IB model for the Suboff case at a higher Reynolds number}

\begin{figure}[!ht]
\centering
\begin{subfigure}[b]{0.48\textwidth}
	\centering
	\includegraphics[width = 1.0\textwidth]{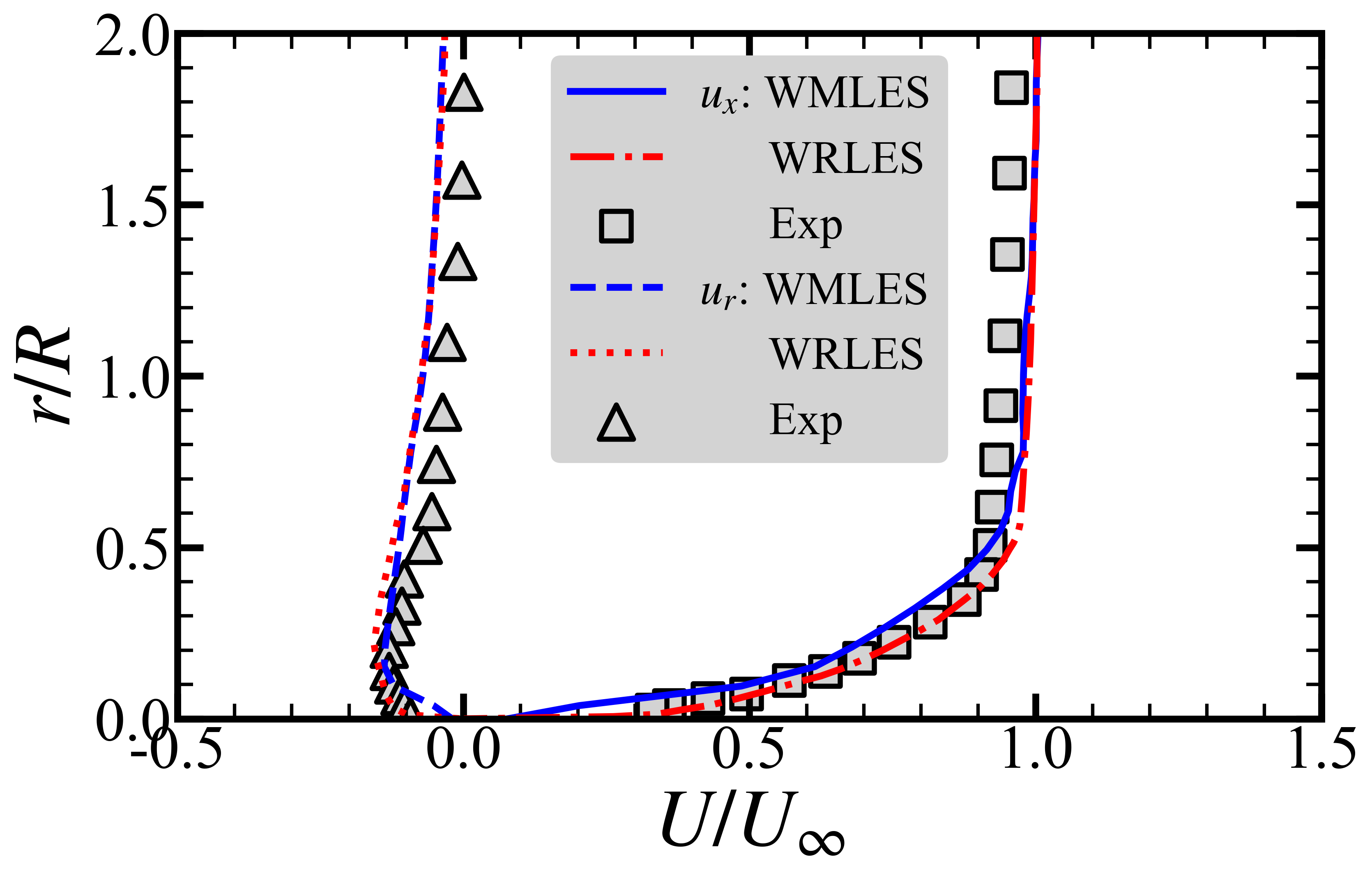}
	\subcaption{$x/L$ = 0.904}
\end{subfigure}
\begin{subfigure}[b]{0.48\textwidth}
	\centering
	\includegraphics[width = 1.0\textwidth]{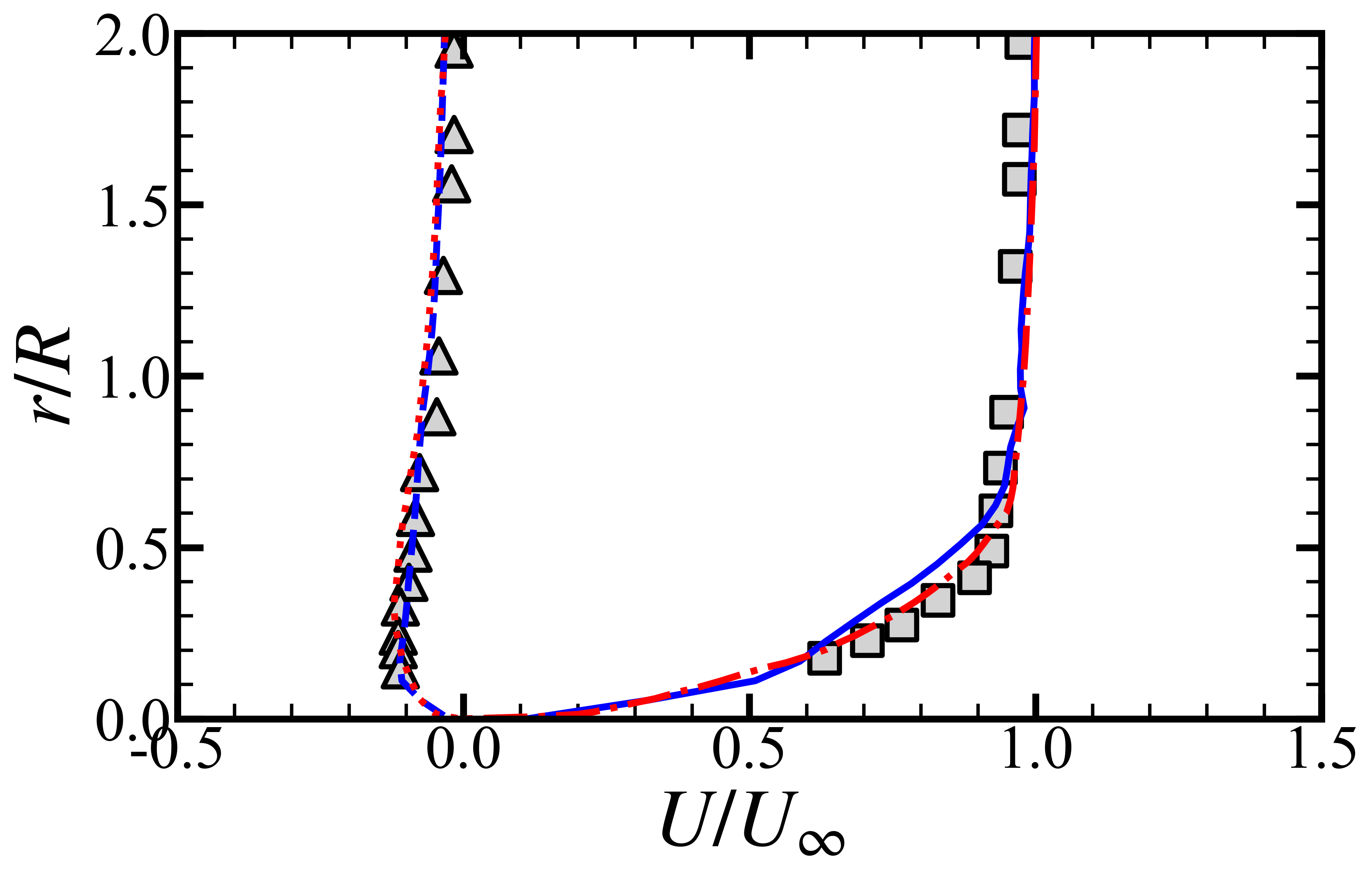}
	\subcaption{$x/L$ = 0.927}
\end{subfigure}
\begin{subfigure}[b]{0.48\textwidth}
	\centering
	\includegraphics[width = 1.0\textwidth]{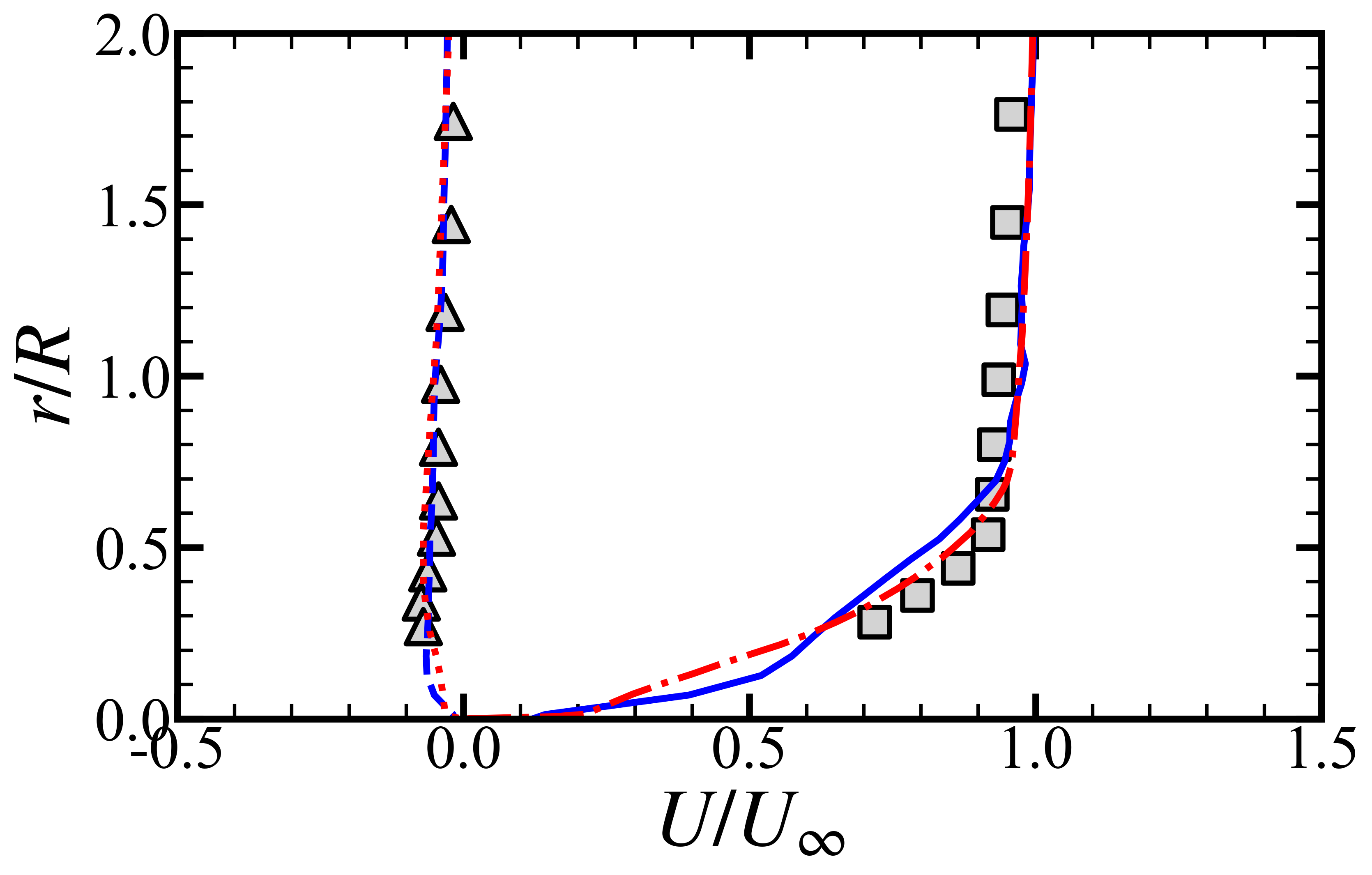}
	\subcaption{$x/L$ = 0.956}
\end{subfigure}
\begin{subfigure}[b]{0.48\textwidth}
	\centering
	\includegraphics[width = 1.0\textwidth]{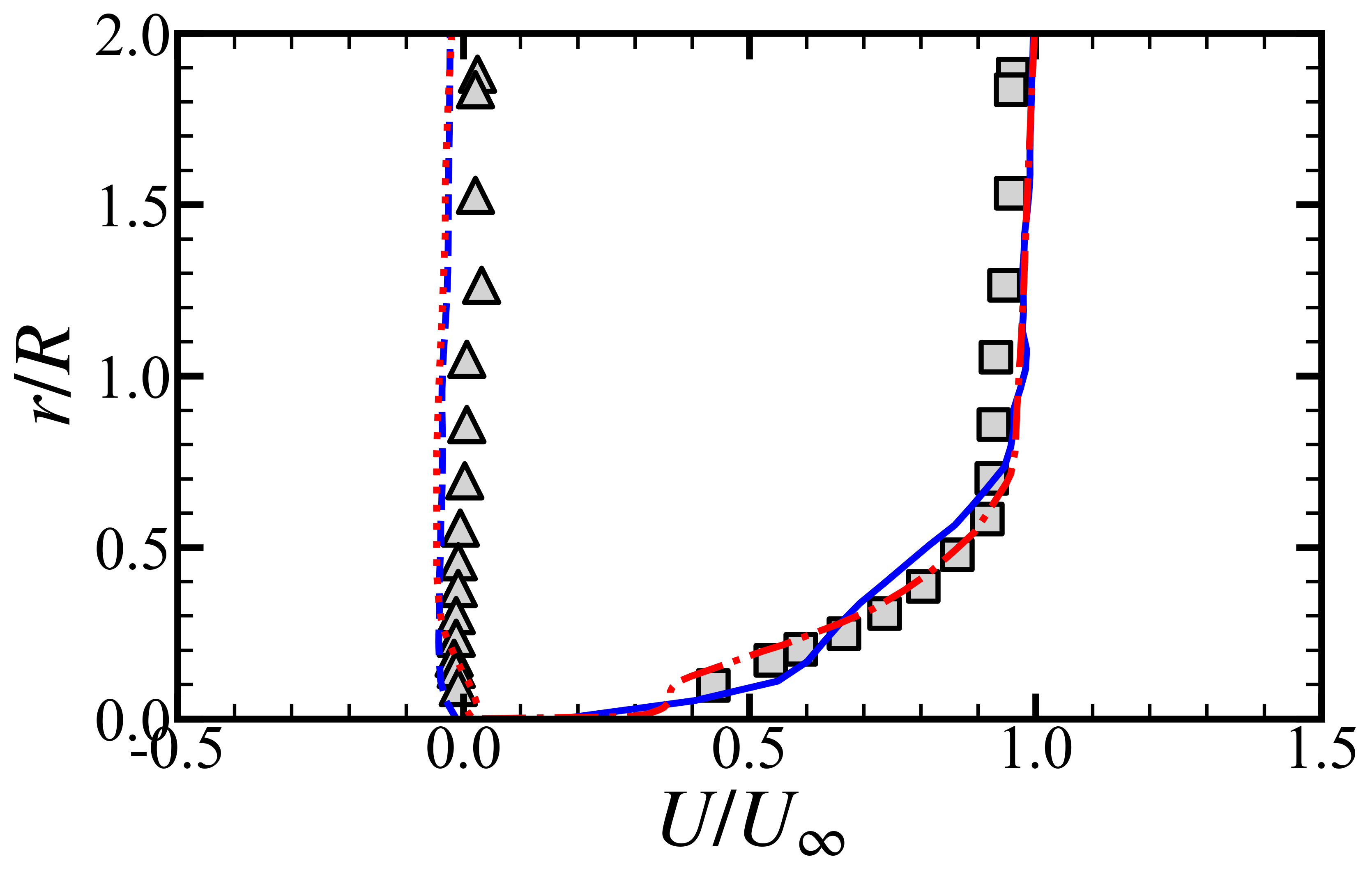}
	\subcaption{$x/L$ = 0.978}
\end{subfigure}
\caption{Comparison of time-averaged streamwise and radial velocity profiles for the Suboff case at $Re_L=1.2\times10^7$ obtained from the present WMLES with G1 grid, the WRLES of Posa \& Balaras~\cite{Posa_Balaras_JFM_2020} and the experiment of Huang \textit{et al.}~\cite{Huang_1994_measurements} at different streamwise locations.}
\label{fig:velo_avg_G1_Re}
\end{figure}
To evaluate the generalization ability of the proposed FEL-IB model, we carry out the WMLES case of flow over the Suboff at a higher Reynolds number ($Re_L=1.2\times 10^7$) using the G1 grid. Figure~\ref{fig:velo_avg_G1_Re} compares the time-averaged axial and radial velocity profiles at different streamwise locations from the present WMLES, the WRLES of Posa \& Balaras~\cite{Posa_Balaras_JFM_2020} and the experiment of Huang \textit{et al.}~\cite{Huang_1994_measurements}. Similar to their previous work~\cite{Posa_Balaras_JFM_2016}, the WRLES data of Posa \& Balaras~\cite{Posa_Balaras_JFM_2020} is obtained from the 2D slice $y+z=0.0$ of the Suboff model with appendages at $Re_L=1.2\times 10^7$. As seen, the axial velocity profiles from the WMLES are consistent with those from the WRLES and experiment at $x/L=0.904 \sim 0.956$, and exhibit a little deviation in the near-wall region at $x/L=0.978$. For the radial velocity profiles, the WMLES result basically agrees with the WRLES result, and both of them exhibit some deviation with the experimental results at $x/L$ = 0.904 and 0.978. Table~\ref{tab:Err_U_G1_Re} lists the relative errors of time-averaged axial velocity between the WMLES with G1 grid and the WRLES/experiment for the Suboff case at $Re_L=1.2\times10^7$. The present WMLES shows an error smaller than 5\% and 3\% at the four streamwise locations compared to the experiment and the WRLES, respectively.
\begin{table}[!ht]
\centering
\caption{\label{tab:Err_U_G1_Re}.The relative error of time-averaged axial velocity between the WMLES with G1 grid and the WRLES or experiment for the Suboff case at $Re_L=1.2\times10^7$.}
\begin{tabular}{p{1.6cm}<{\centering}p{2.5cm}<{\centering}p{2.5cm}<{\centering}}
  \hline
  $x/L$ & $Err_U$-exp & $Err_U$-WRLES  \\
  \hline
  0.904  &  4.93\%  &  2.79\%  \\
  0.927  &  3.46\%  &  1.37\%  \\
  0.956  &  4.60\%  &  1.29\%  \\
  0.978  &  4.52\%  &  2.34\%  \\
  \hline
\end{tabular}
\end{table}
\begin{figure}[!ht]
\centering
\begin{subfigure}[b]{0.48\textwidth}
	\centering
	\includegraphics[width = 1.0\textwidth]{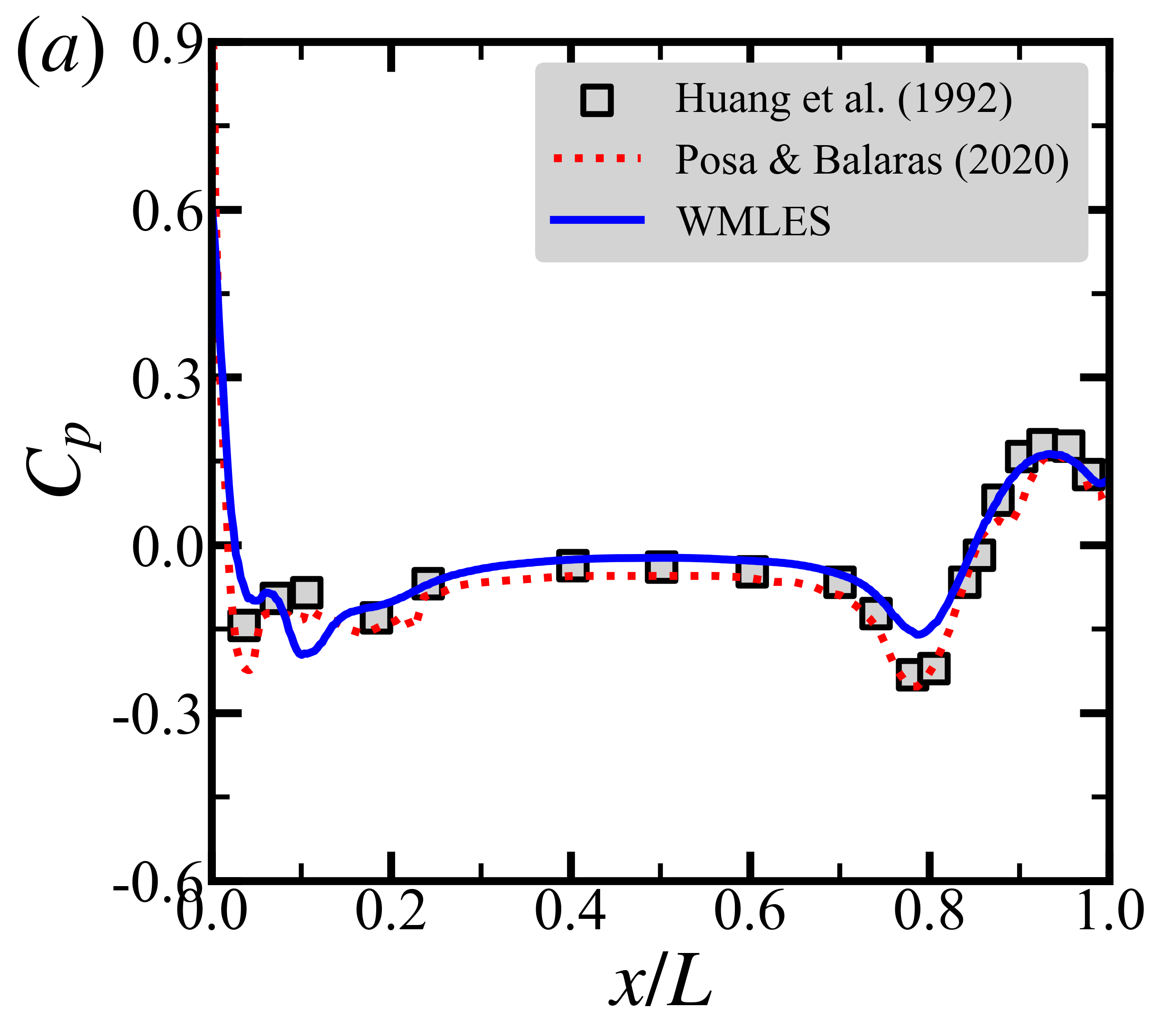}
\end{subfigure}
\begin{subfigure}[b]{0.48\textwidth}
	\centering
	\includegraphics[width = 1.0\textwidth]{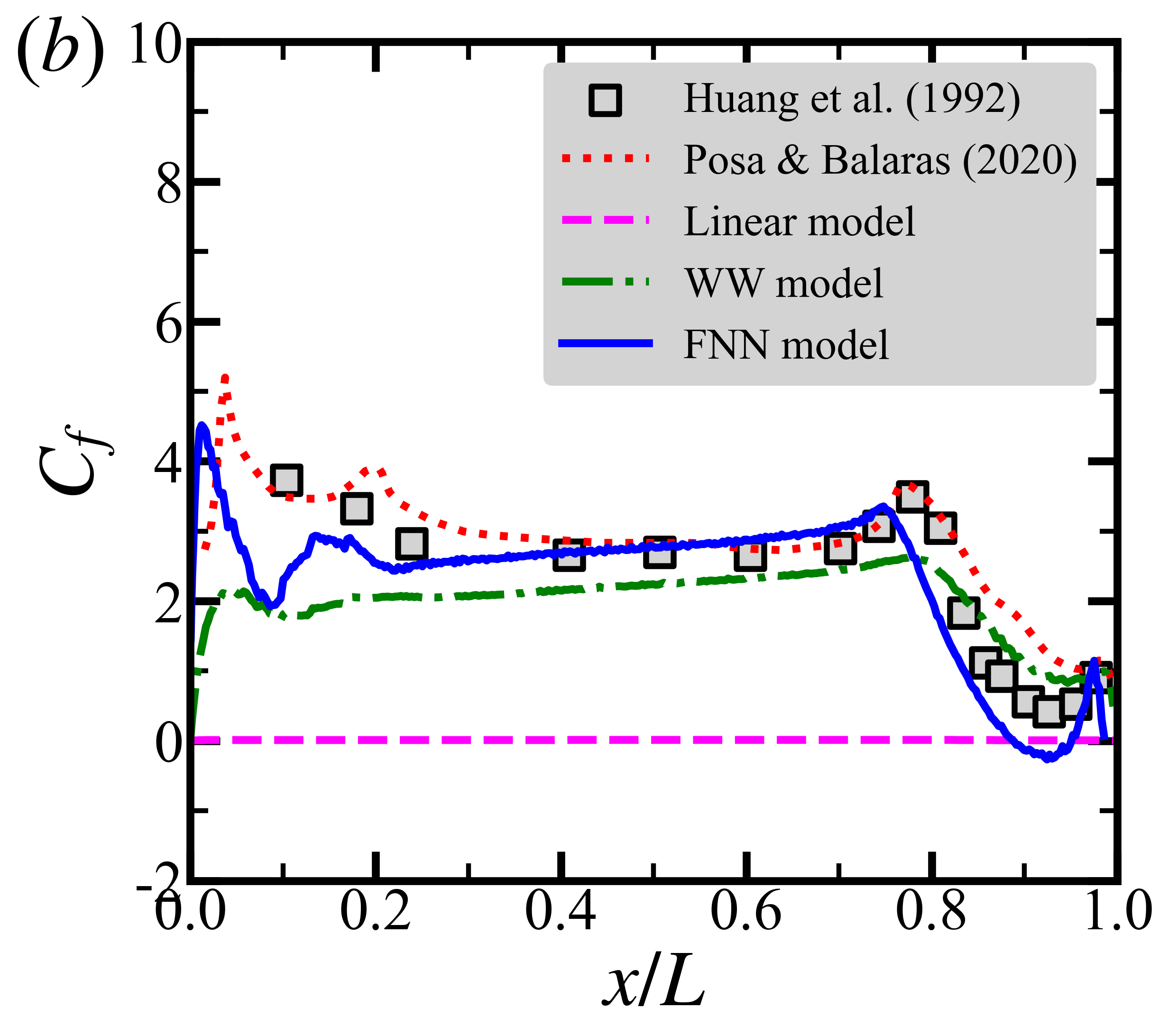}
\end{subfigure}
\caption{Comparison of pressure coefficient $C_p$ and skin-friction coefficient $C_f$ for the Suboff case at $Re_L=1.2\times10^7$ obtained from the present WMLES with G1 grid, the WRLES of Posa \& Balaras~\cite{Posa_Balaras_JFM_2020} and the experiment of Huang \textit{et al.}~\cite{Huang_1994_measurements}.}
\label{fig:Cp_Cf_G1_Re}
\end{figure}
\begin{figure}[!ht]
\centering
\begin{subfigure}[b]{0.48\textwidth}
	\centering
	\includegraphics[width = 1.0\textwidth]{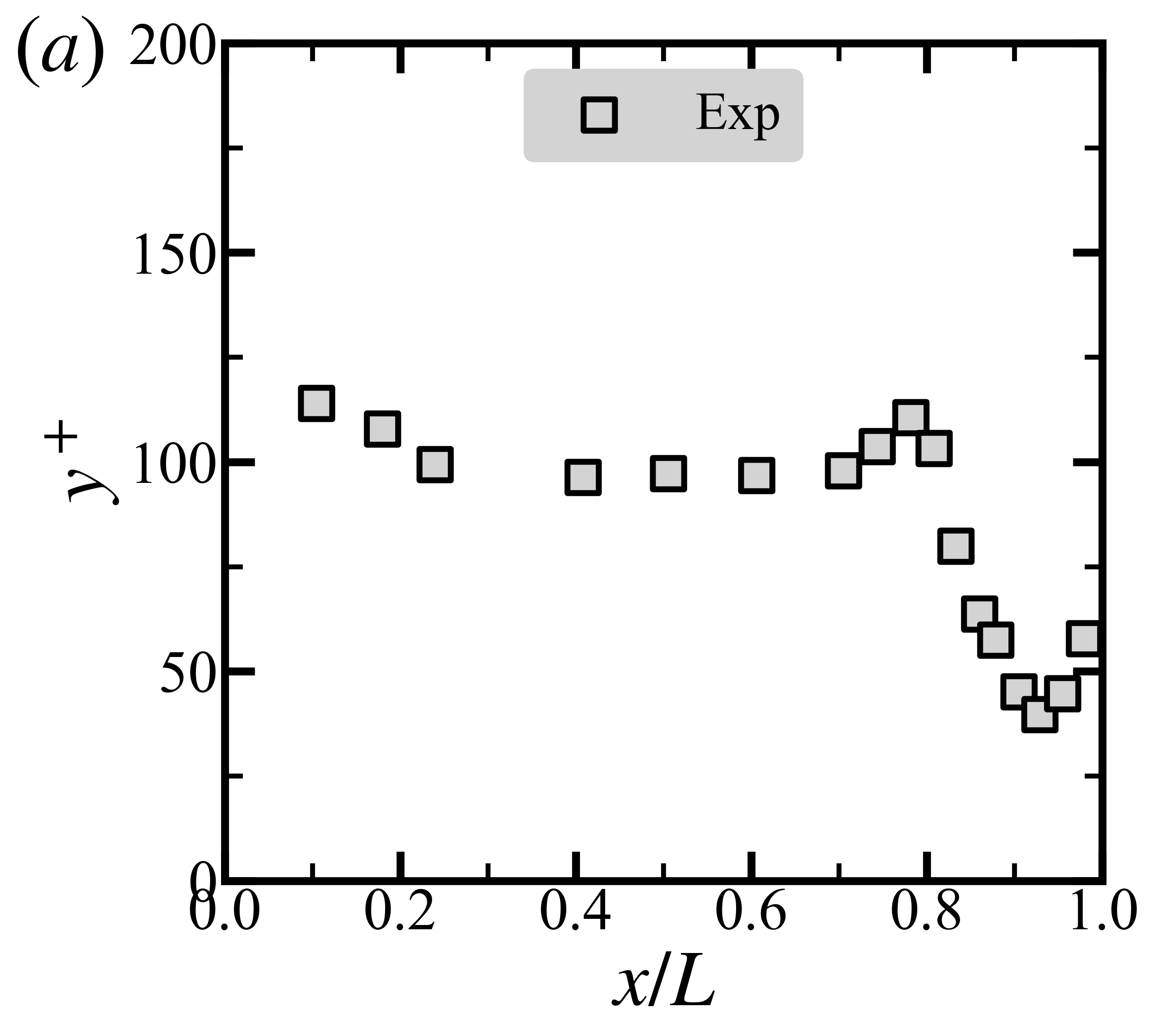}
\end{subfigure}
\begin{subfigure}[b]{0.48\textwidth}
	\centering
	\includegraphics[width = 1.0\textwidth]{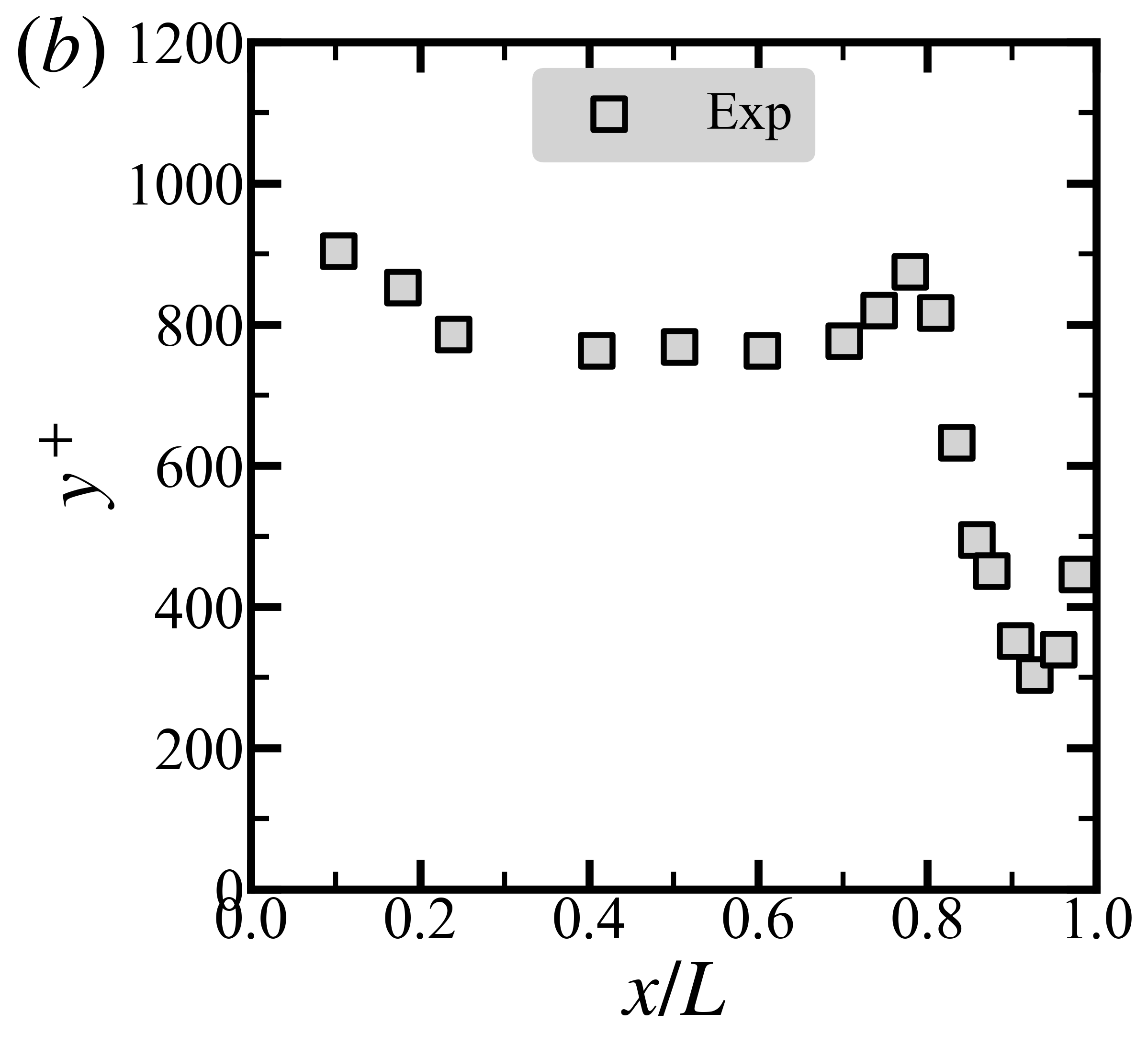}
\end{subfigure}
\caption{The nondimensional wall-normal distance $y^+$ versus the streamwise location $x/L$ for the Suboff case with G1 grid at (a) $Re_L=1.2\times10^6$ and (b) $Re_L=1.2\times10^7$. The values are estimated using the skin-friction coefficient from the experiment of Huang \textit{et al.}~\cite{Huang_1994_measurements}.}
\label{fig:G1_yplus}
\end{figure}

Figure~\ref{fig:Cp_Cf_G1_Re} compares the pressure and skin-friction coefficients obtained from the present WMLES, the WRLES of Posa \& Balaras~\cite{Posa_Balaras_JFM_2020} and the experiment of Huang \textit{et al.}~\cite{Huang_1994_measurements} for the Suboff case at $Re_L=1.2\times10^7$. In figure~\ref{fig:Cp_Cf_G1_Re}(a), the pressure coefficient from the WMLES basically agrees with the different referenced results, with slight underestimation at the beginning of the stern with contraction ($x/L=0.78$). For the skin-friction coefficient in figure~\ref{fig:Cp_Cf_G1_Re}(b),
the curves from the WRLES and the experiment are very consistent at $x/L \le 0.8$, but exhibit significant deviations at the stern ($x/L > 0.8$). The WMLES case with G1 grid provides three different predictions of $C_f$: the linear model gives poor prediction at all streamwise locations, the WW model only accurately predicts $C_f$ at the parallel middle body, and the FNN model~\cite{Zhou_etal_PoF_2023} accurately predicts $C_f$ at most of streamwise locations, especially at the stern with adverse pressure gradient. 

To quantify the thickness of grid resolution, the nondimensional distance $y^+ = y_G u_\tau/ 2\nu$ of the first off-wall grid for the Suboff case with G1 grid is estimated using the skin-friction coefficient from the experiment of Huang \textit{et al.}~\cite{Huang_1994_measurements} and the transverse grid size $y_G = 0.03D$ at the centre region. Figure~\ref{fig:G1_yplus} shows the nondimensional distance $y^+$ versus the streamwise location $x/L$ for the Suboff case at different Reynolds numbers. At the middle body, the value of $y^+$ is approximately 100 and 800 at $Re_L=1.2\times10^6$ and $1.2\times10^7$, respectively. As for the stern of the Suboff, the value of $y^+$ decreases to 50 and 400, respectively.

Therefore, the proposed FEL-IB model is successful in predicting the time-averaged velocity profiles, the pressure and skin-friction coefficients for the flow over the Suboff at a higher Reynolds number than the training case using the G1 coarse grid.

\subsubsection{Evaluation on the velocity fluctuation}

\begin{figure}[!ht]
\centering
\begin{subfigure}[b]{0.48\textwidth}
	\centering
	\includegraphics[width = 1.0\textwidth]{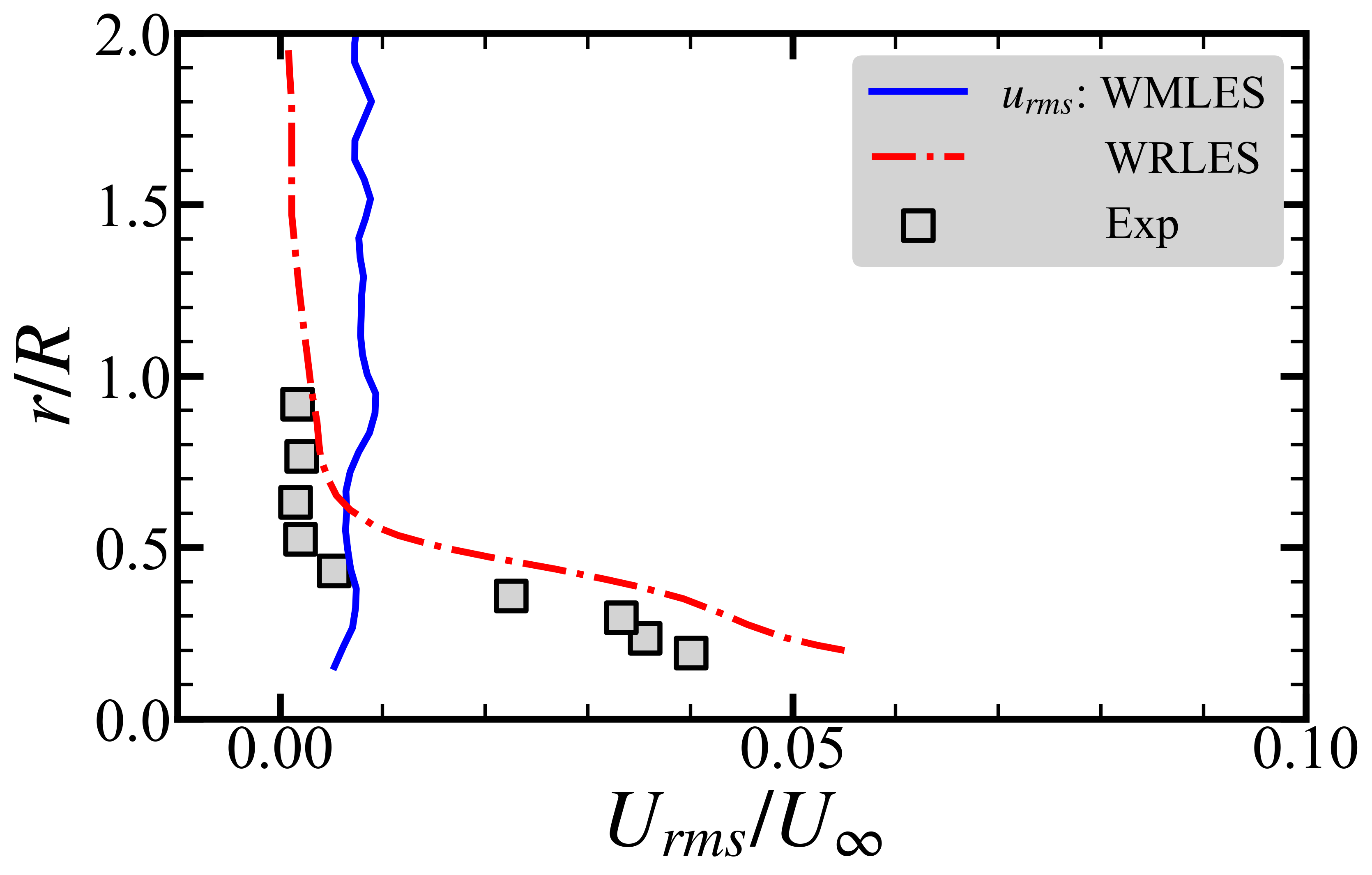}
	\subcaption{$x/L$ = 0.904}
\end{subfigure}
\begin{subfigure}[b]{0.48\textwidth}
	\centering
	\includegraphics[width = 1.0\textwidth]{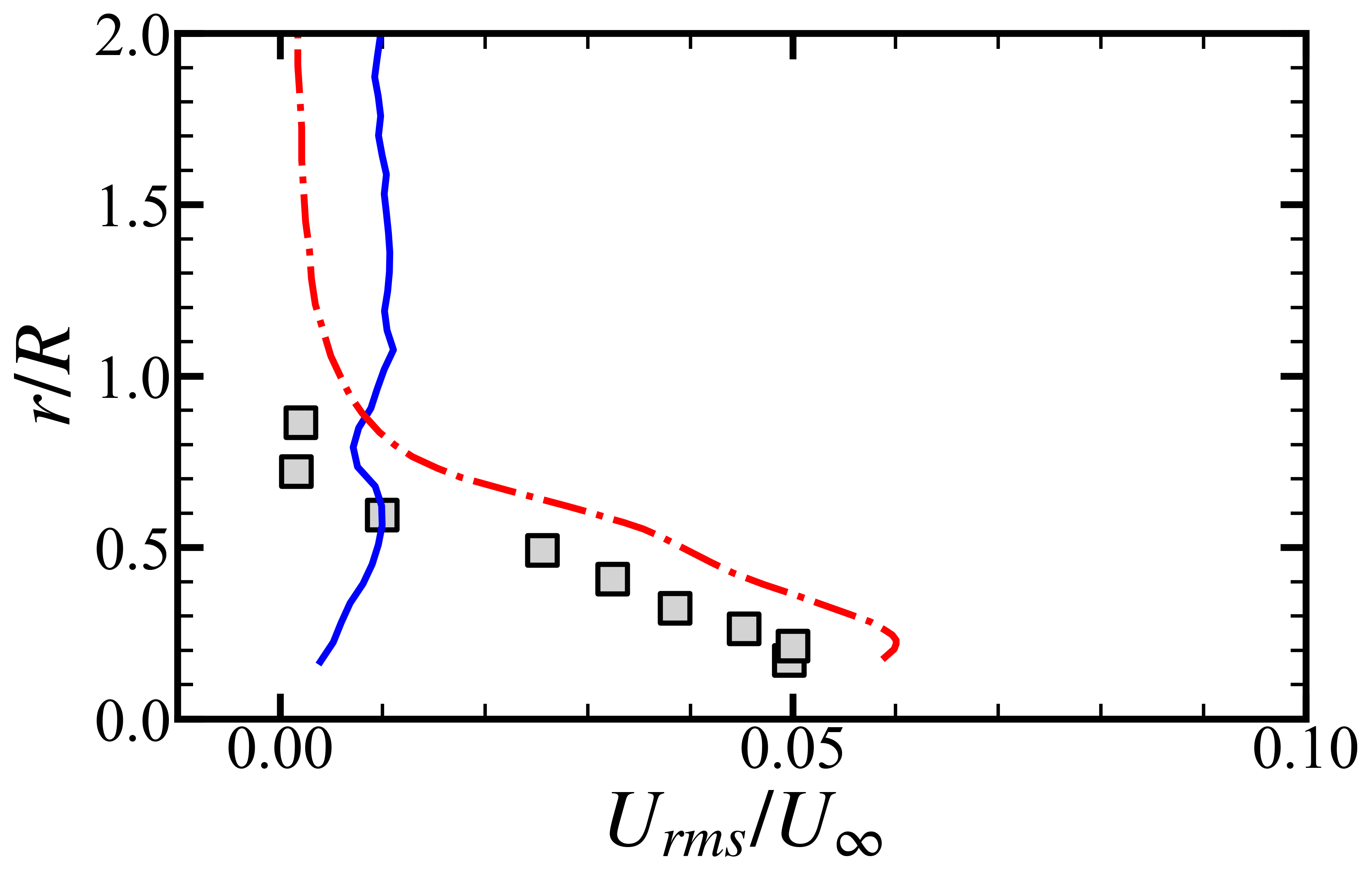}
	\subcaption{$x/L$ = 0.978}
\end{subfigure}
\caption{Comparison of the rms of axial velocity profiles for the Suboff case at $Re_L=1.2\times10^6$ obtained from the present WMLES with G1 grid, the WRLES of Kumar \& Mahesh~\cite{Kumar_Mahesh_JFM_2018} and the experiment of Huang \textit{et al.}~\cite{Huang_1994_measurements}.}
\label{fig:velo_rms_G1}
\end{figure}
\begin{figure}[!ht]
\centering
\begin{subfigure}[b]{0.48\textwidth}
	\centering
	\includegraphics[width = 1.0\textwidth]{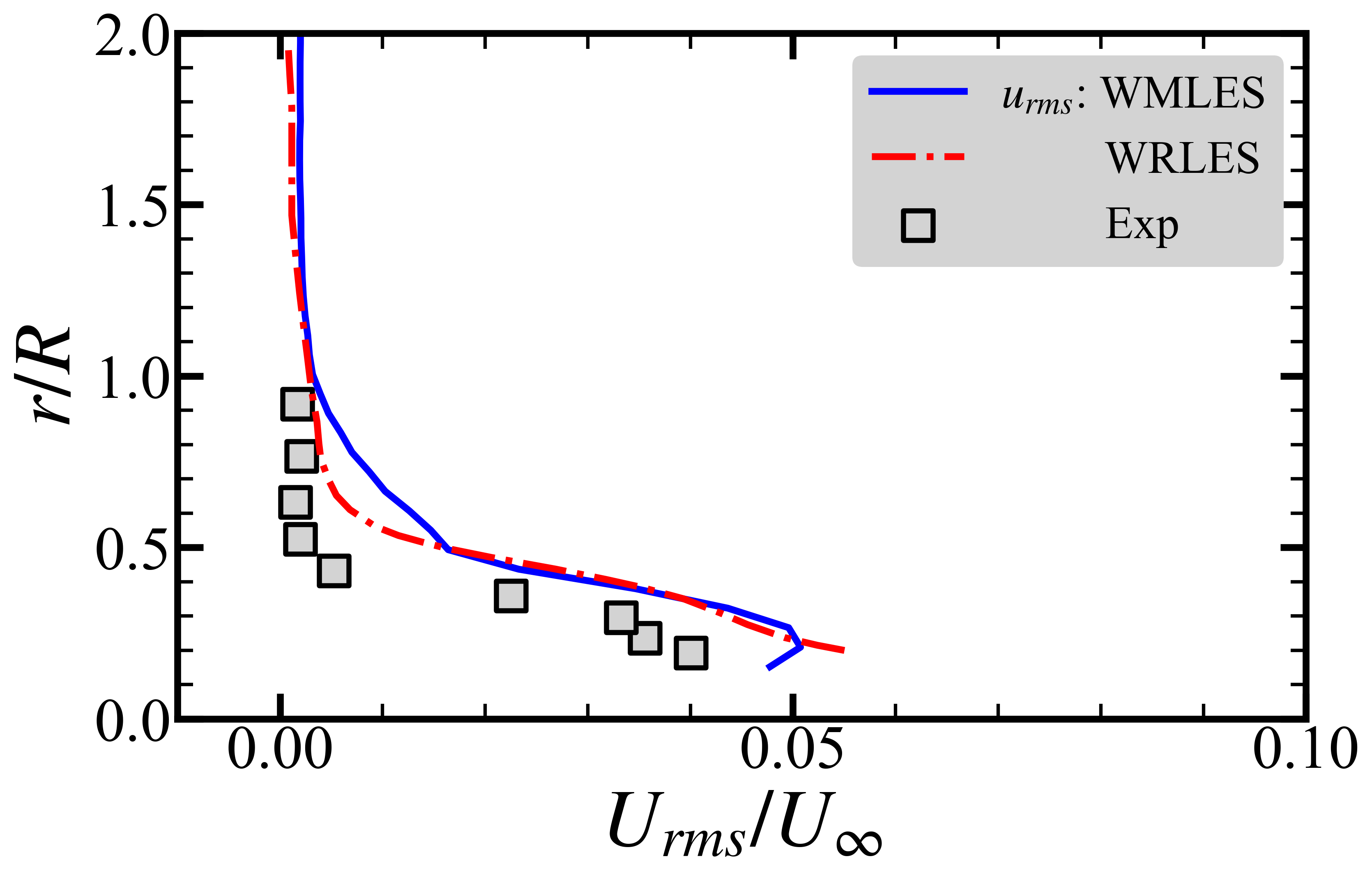}
	\subcaption{$x/L$ = 0.904}
\end{subfigure}
\begin{subfigure}[b]{0.48\textwidth}
	\centering
	\includegraphics[width = 1.0\textwidth]{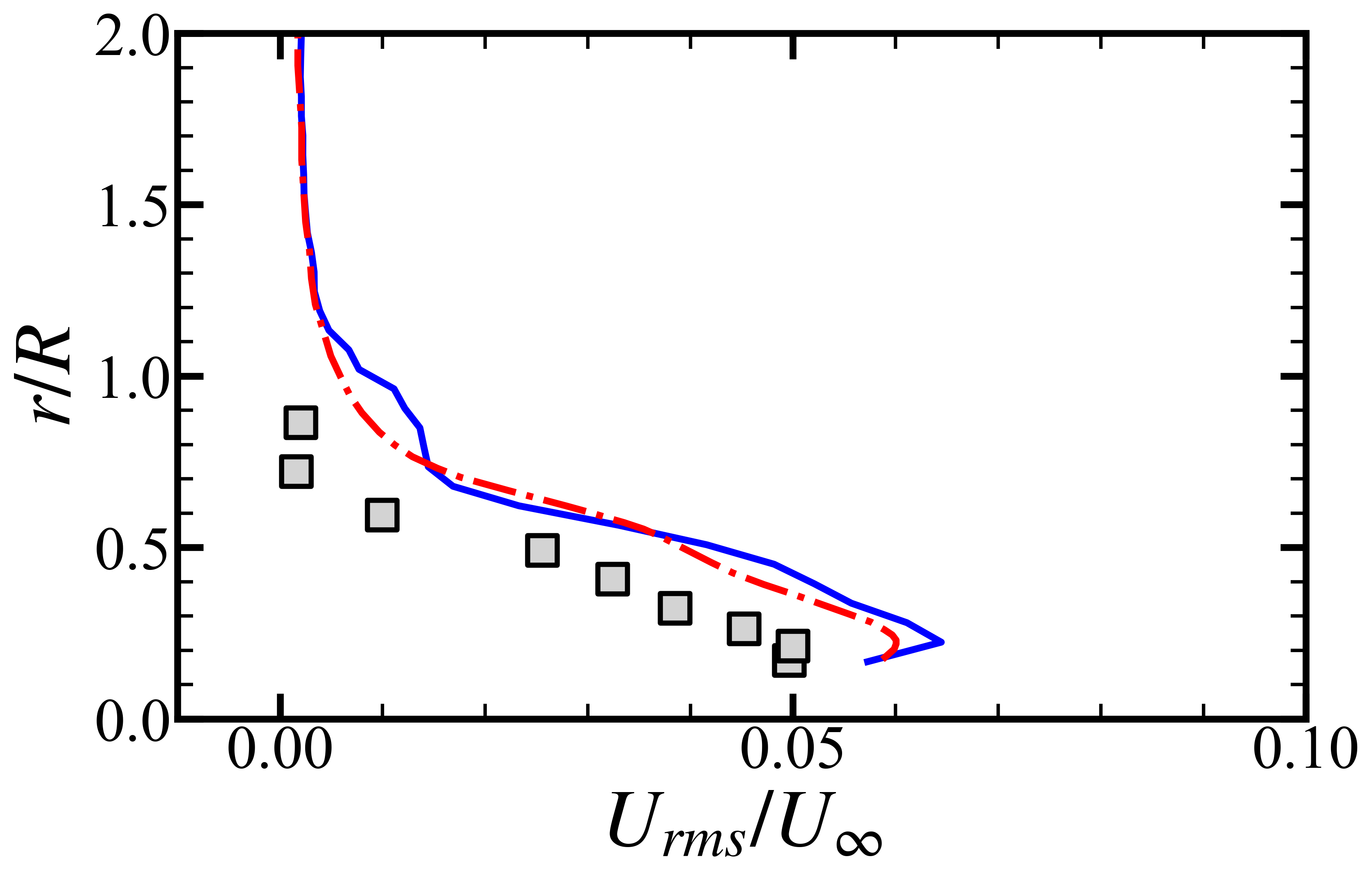}
	\subcaption{$x/L$ = 0.978}
\end{subfigure}
\caption{Comparison of the rms of axial velocity profiles for the Suboff case at $Re_L=1.2\times10^6$ obtained from the present WMLES with G2 grid, the WRLES of Kumar \& Mahesh~\cite{Kumar_Mahesh_JFM_2018} and the experiment of Huang \textit{et al.}~\cite{Huang_1994_measurements}.}
\label{fig:velo_rms_G2}
\end{figure}
\begin{table}[!ht]
\centering
\caption{\label{tab:Err_U_G2}.The relative error of time-averaged axial velocity and the rms of axial velocity between the WMLES with G2 grid and the WRLES or experiment for the Suboff case at $Re_L=1.2\times10^6$.}
\begin{tabular}{p{1.4cm}<{\centering}p{2.2cm}<{\centering}p{2.3cm}<{\centering}p{3.0cm}<{\centering}p{3.5cm}<{\centering}}
  \hline
  $x/L$  &  G2: $Err_U$-exp  &  $Err_U$-WRLES  & $Err_{U, \text{rms}}$-WRLES & G1: $Err_{U, \text{rms}}$-WRLES  \\
  \hline
  0.904  &  4.72\%  & 1.12\%  &  12.96\%  &  77.29\%  \\
  0.927  &  2.58\%  & --      &  --       &  --       \\
  0.956  &  3.90\%  & --      &  --       &  --       \\
  0.978  &  5.34\%  & 1.73\%  &  6.79\%   &  78.13\%  \\
  \hline
\end{tabular}
\end{table}

To further examine the prediction of velocity fluctuation, the WMLES of flow over the Suboff at $Re_L=1.2\times10^6$ is carried out using the G2 fine grid.
Figures~\ref{fig:velo_rms_G1} and \ref{fig:velo_rms_G2} show the profiles of root mean square (rms) of the axial velocity $U_{rms}$ obtained from the present WMLES of Suboff case with the G1 and G2 grid, respectively, while the WRLES data of Kumar \& Mahesh~\cite{Kumar_Mahesh_JFM_2018} and the experimental data of Huang \textit{et al.}~\cite{Huang_1994_measurements} are also included as comparison. There are significant deviations of the rms velocity between the WMLES and WRLES using the G1 coarse grid. Once the G2 fine grid is used, the rms velocity profiles predicted by the WMLES coincide well with those from the WRLES, but both of them exhibit the larger values than the experimental data.

The relative errors of the time-averaged axial velocity ($Err_U$) and the rms of axial velocity ($U_{\text{rms}}$) between the WMLES with G2 grid and the WRLES or experiment are then listed in table~\ref{tab:Err_U_G2}. Using the refined G2 grid, the values of $Err_U$ are close to those from the G1 grid, demonstrating that the time-averaged velocity profiles are accurately predicted (not shown for the sake of brevity). As for the rms of axial velocity, the error between the WMLES and WRLES is 12.96\% at $x/L=0.904$ and 6.79\% at $x/L=0.978$, which are significantly smaller than those from the G1 grid.



\section{Conclusions}\label{sec:Conclusion}
A features-embedded-learning-immersed boundary (FEL-IB) wall model is proposed to approximate immersed boundary (IB) conditions in the hybrid wall-modeled large-eddy simulation (WMLES) and IB framework for simulating high-Reynolds-number wall-bounded turbulent flows.
In the FEL-IB wall model, a neural network trained within the WMLES flow solver using the ensemble Kalman method estimates the momentum flux at the IB-fluid interface based on flow quantities at three off-wall positions, thereby providing boundary conditions for the viscous term. For the advection term, the velocity at IB nodes—determined using the classical WW model—serves as the boundary condition.

To evaluate the model's performance, we simulate two challenging test cases: the flow over a body of revolution and flow around the DARPA Suboff model. For the body of revolution at $Re_L=1.9 \times 10^6$, the FEL-IB model accurately predicts vertical profiles of time-averaged velocity at multiple streamwise locations. 
For the Suboff simulations, WMLES cases with two grid resolutions and Reynolds numbers ($Re_L=1.2 \times 10^6$ and $1.2 \times 10^7$) were examined. At the lower Reynolds number ($Re_L=1.2 \times 10^6$), the WMLES with FEL-IB achieves: 1) accurate prediction of mean velocity profiles on coarse grids, and 2) improved resolution of root-mean-square (RMS) axial velocity fluctuations on finer grids.
Notably, the FEL-IB shear stress model—pretrained using periodic hill flow data and the logarithmic law—outperforms both wall-resolved LES (WRLES) and classical WW-model WMLES in predicting skin-friction coefficients, particularly at the Suboff's stern region where the adverse pressure gradients dominate. At the higher Reynolds number ($Re_L=1.2 \times 10^7$), the model maintains good agreement for the time-averaged velocity profiles, and the pressure and skin-friction coefficient distributions. 

\section*{Acknowledgment}
This work was supported by NSFC Basic Science Center Program for ``Multiscale Problems in Nonlinear Mechanics'' (NO. 11988102), the Strategic Priority Research Program of Chinese Academy of Sciences (CAS, NO. XDB0620102), National Natural Science Foundation of China (NO. 12172360), and CAS Project for Young Scientists in Basic Research (YSBR-087).

\bibliography{ref}

\end{document}